\documentstyle{article}
\def\fcapslashcD{{\cal D}\!\!\!\!/\,}
\def\fcapslashD{D\!\!\!\!/\,}
\newenvironment{keylistone}%
 {%
 \begin{list}{}%
 {%
 \setlength{\itemindent}{0pt}%
 \setlength{\itemsep}{\baselineskip}%
 \setlength{\labelsep}{15pt}%
 \settowidth{\labelwidth}{{Table 9.}}%
 \setlength{\leftmargin}{\labelwidth}%
 \addtolength{\leftmargin}{\labelsep}%
 \addtolength{\leftmargin}{10pt}%
 \setlength{\rightmargin}{15pt}%
 }%
 }%
 { \end{list} }
\begin{document}
\rightline{UG-9/92}
\vskip 1in
\centerline{\large\bf The Supersymmetric Effective Action of the}
\vskip 12pt
\centerline{\large\bf Heterotic String in Ten Dimensions}
\vskip 50pt
\centerline{M. de Roo, H. Suelmann and A. Wiedemann}
\vskip 12pt
\centerline{\sl Institute for Theoretical Physics}
\centerline{\sl University of Groningen}
\centerline{\sl Nijenborgh 4, 9747 AG Groningen}
\centerline{\sl The Netherlands}
\vfill
{\bf Abstract}
\vskip 7pt
We construct the supersymmetric completion of
 quartic $R+R^4$-actions in the ten-dimensional effective action of
 the heterotic string. Two
 invariants, of which the bosonic parts are known from one-loop
 string
 amplitude calculations, are obtained. One of these
 invariants can be generalized to an $R+F^2+F^4$-invariant for
 supersymmetric Yang-Mills theory coupled to supergravity.
 Supersymmetry requires the presence of $B\wedge R\wedge R\wedge
 R\wedge R$-terms, ($B\wedge F\wedge F\wedge F\wedge F$ for
 Yang-Mills) which correspond to counterterms in the Green-Schwarz
 anomaly cancellation. Within the context of our calculation
 the $\zeta(3)R^4$-term from the tree-level string effective action
 does not allow supersymmetrization.
\par
\vskip 50 pt
\noindent
Groningen, October 1992

\newpage
\section{$N=1$ supergravity in $d=10$ and the $R^2$-action.}
\vskip 12pt
The basic multiplet of $N=1$ supergravity in ten
 spacetime dimensions consists of the tenbein field $e_\mu{}^a$,
 the dilaton field $\phi$, an antisymmetric tensor gauge
 field $B_{\mu\nu}$ and the Majorana-Weyl fermions $\psi_\mu$
 (gravitino)
 and $\lambda$ (dilatino) \cite{Cham81}.
 This multiplet transforms under local
 supersymmetry as follows\footnote{
 Note that $\Gamma^{a_1\ldots a_n}
 \equiv \Gamma^{[a_1}\Gamma^{a_2}\ldots \Gamma^{a_n]}$.
 Throughout this paper we use the
 conventions of \cite{BedR89}. In our calculations we will never
 consider terms quartic in fermions in the action, and,
 consequently, we may ignore terms quadratic in fermions in the
 transformation rules.}:
\begin{eqnarray}
\label{2.1}
 \delta e_\mu{}^a & = & {\textstyle{1\over 2}}\bar\epsilon\,
 \Gamma^a\psi_\mu \,, \\
\label{2.2}
 \delta\psi_\mu & = &
 (\partial_\mu-{\textstyle{1\over 4}}\Omega_{\mu}{}^{ab}_+
 \Gamma_{ab})\,\epsilon+\epsilon\, ({\rm fermi})^2 \,, \\
\label{2.3}
 \delta B_{\mu\nu} & = & {\textstyle{1\over 2}}\sqrt{2}\,
 \bar\epsilon\,\Gamma_{[\mu}\psi_{\nu]} \,, \\
\label{2.4}
 \delta\lambda & = & -{\textstyle{3\over 8}}\sqrt{2}\,
 \Gamma^\mu\epsilon\,\phi^{-1}{D}_\mu\phi
 +{\textstyle{1\over 8}}\Gamma^{abc}\epsilon\,
 \hat H_{abc}
 +\epsilon\,({\rm fermi})^2 \,, \\
\label{2.5}
 \phi^{-1}\delta\phi & = & -{\textstyle{1\over 3}}\sqrt{2}\,
 \bar\epsilon\lambda \,.
\end{eqnarray}

The derivatives ${\cal D}_\mu$ are Lorentz covariant,
 supercovariant derivatives are denoted by $D_\mu$. In the
 variation of the gravitino field we encounter a torsionful spin
 connection defined by:
\begin{eqnarray}
\label{2.6}
 \Omega_{\mu}{}^{ab}_\pm & \equiv & \omega_\mu{}^{ab}(e,\psi)
 \pm{\textstyle{3\over 2}}\sqrt{2}\,
 \hat{H}_\mu{}^{ab} \,.
\end{eqnarray}
Here, $\omega_\mu{}^{ab}(e,\psi)$ is the usual spin connection
 with $\psi$-torsion, i.e., the solution
 of $D_{[\mu}(\omega)e_{\nu]}{}^a=0$. The additional torsion is
 determined by the supercovariant field strength of
 the $B$-field,
 $\hat{H}_\mu{}^{ab}$, given by:
\begin{eqnarray}
\label{2.7}
\hat{H}_{\mu\nu\rho} & = & \partial_{[\mu}B_{\nu\rho]}
 -{\textstyle{1\over 4}}
 \bar\psi_{[\mu}\Gamma_\nu\psi_{\rho]} \,,
\end{eqnarray}
which is invariant under gauge transformations
\begin{eqnarray}
\label{2.8}
\delta B_{\mu\nu} & = &
 \partial_\mu\Lambda_\nu-\partial_\nu\Lambda_\mu \,.
\end{eqnarray}
Under local supersymmetry, $\omega_\mu{}^{ab}$ and $\hat H$
 transform as
\begin{eqnarray}
\label{2.9}
\delta\omega_\mu{}^{ab}(e,\psi)
 & = &{\textstyle{1\over 4}}\bar\epsilon\,
 \Gamma_\mu\psi^{ab}
 +{\textstyle{1\over 2}}\bar\epsilon\,
 \Gamma^{[a}\psi_\mu{}^{b]}
 +{\textstyle{3\over 4}}\sqrt{2}\,\bar\epsilon\,
 \Gamma_c\psi_\mu
 \,\hat{H}^{abc} \,, \\
\label{2.11}
\delta\hat{H}_{abc} & = & -{\textstyle{1\over 4}}\sqrt{2}\,
 \bar\epsilon\,\Gamma_{[a}\psi_{bc]} \,.
\end{eqnarray}
Here, $\psi_{ab}$ denotes the gravitino curvature
\begin{eqnarray}
\label{2.10}
 \psi_{\mu\nu} & = & 2\,{\cal D}_{[\mu}(\Omega_+)\psi_{\nu]}
 +({\rm fermi})^3 \,.
\end{eqnarray}

\par
The transformations (\ref{2.9}) and (\ref{2.11}) can be combined to
 yield:
\begin{eqnarray}
\label{2.12}
 \delta\Omega_{\mu}{}^{ab}_-
 & = & {\textstyle{1\over 2}}\bar\epsilon\,
 \Gamma_\mu\psi^{ab} \,.
\end{eqnarray}
The gravitino curvature $\psi^{ab}$ itself has the following
 variation:
\begin{eqnarray}
\label{2.13}
\delta\psi^{ab} & = & -{\textstyle{1\over 4}}
 \Gamma^{\mu\nu}\epsilon\, R_{\mu\nu}{}^{ab}(\Omega_-)
 +\epsilon\,({\rm fermi})^2 \,,
\end{eqnarray}
where $R_{\mu\nu}{}^{ab}(\Omega_-)$ denotes the Riemann curvature
 tensor with spin connection $\Omega_-$.
\par
The ten-dimensional action which is invariant under the
 transformations (\ref{2.1}--\ref{2.5}) is given by:
\begin{eqnarray}
\label{2.14}
 {\cal L}_R & = & e\phi^{-3}\Bigl\{
 -{\textstyle{1\over 2}}R(\omega(e))
 -{\textstyle{3\over 4}}H_{\mu\nu\lambda}H^{\mu\nu\lambda}
 +{\textstyle{9\over 2}}(\phi^{-1}\partial_\mu\phi)^2
 \nonumber \\
 & &\phantom{e2224}
 -{\textstyle{1\over 2}}\bar\psi_\mu\Gamma^{\mu\nu\rho}
 {\cal D}_\nu(\omega(e))\psi_\rho
 +2\sqrt{2}\,\bar\lambda\Gamma^{\mu\nu}
 {\cal D}_\mu(\omega(e))\psi_\nu
 \nonumber \\
 & &\phantom{e2224}
 +4\bar\lambda
 \fcapslashcD (\omega(e))\lambda
 +3\sqrt{2}\,\bar\psi_\mu\Gamma^\nu\Gamma^\mu\lambda
 (\phi^{-1}\partial_\nu\phi)
 \nonumber \\
 & &\phantom{e2224}
 - {\textstyle{3\over 2}}\bar\psi_\mu\Gamma^\mu\psi_\nu
 (\phi^{-1}\partial^\nu\phi)
 \nonumber \\
 & &\phantom{e2224}
 +{\textstyle{1\over 16}}\sqrt{2}\,H^{\rho\sigma\tau}
 \bigl[\bar\psi_\mu\Gamma^{[\mu}\Gamma_{\rho\sigma\tau}
 \Gamma^{\nu]}\psi_\nu
 +4\sqrt{2}\,\bar\psi^\mu\Gamma_{\mu\rho\sigma\tau}
 \lambda \nonumber \\
 & &\phantom{gjhgjhgjhgjhgjhgjhgj}
 -8\bar\lambda\Gamma_{\rho\sigma\tau}\lambda
 \bigr] \Bigr\} +({\rm fermi})^4\,.
\end{eqnarray}
The equations of motion which
 follow from (\ref{2.14}) will play an important role in this paper.
 These equations, for the fields $\phi$, $e_\mu{}^a$,
 $\bar\lambda$, $\bar\psi_\mu$ and $B_{\mu\nu}$, respectively, read:
\begin{eqnarray}
 \Phi & = &e\phi^{-3}\Bigl({\textstyle{3\over 2}}R(\omega)
 -9{\cal D}_a(\omega)(\phi^{-1}\partial^a\phi)
 +{\textstyle{27\over 2}}
 (\phi^{-1}\partial_\mu\phi)^2
 \nonumber \\ \label{2.15}
 &&\qquad
 +{\textstyle{9\over 4}}
 H_{\mu\nu\lambda}H^{\mu\nu\lambda}
 \Bigr)\,, \\
 {\cal E}^\mu{}_a & = &
 e\phi^{-3}e^\mu{}_be^\nu{}_a\Bigl(
 R_\nu{}^b(\omega)
 -3{\cal D}_\nu(\omega)(\phi^{-1}\partial^b\phi)
 +{\textstyle{9\over 2}}H_{\nu\lambda\rho}
 H^{b\lambda\rho}
 \Bigr)
 \nonumber \\
 \label{2.16}&&\qquad
 -{\textstyle{1\over 3}}e^\mu{}_a\Phi\,, \\
\label{2.17}
\Lambda & = & e\phi^{-3}
 \Bigl\{8\fcapslashD (\omega)\lambda
 +\sqrt{2}\,\Gamma^{\mu\nu}\psi_{\mu\nu}
 -12(\phi^{-1}\fcapslashD \phi)\lambda
 -\sqrt{2}\,\Gamma^{abc}\lambda H_{abc}
 \Bigr\}\,, \\
\label{2.18}
\Psi_\mu & = & e\phi^{-3}\Bigl(\Gamma^\nu\psi_{\mu\nu}
 +2\sqrt{2}\,D_\mu(\Omega_+)\lambda\Bigr)
 -{\textstyle{1\over 4}}\sqrt{2}\,\Gamma_\mu\Lambda
 \,, \\
\label{2.19}
{\cal B}^{\mu\nu} & = & {\textstyle{3\over 2}}
 \partial_\lambda\bigl(
 e\phi^{-3}H^{\lambda\mu\nu}\bigr) \,.
\end{eqnarray}
In this paper we frequently use identities
 which are implied by the fermionic equations of motion
 and the Bianchi identity
\begin{eqnarray}
\label{2.10a}
 {\cal D}_{[\mu}(\omega)\psi_{\nu\lambda]}
 & = &-{\textstyle{1\over 4}}\Gamma^{ab}
 \psi_{[\mu}\,R_{\nu\lambda]ab}(\omega) \,
\end{eqnarray}
 for the gravitino curvature.
 First of all, we use (\ref{2.18}) to solve for the single
 $\Gamma$-contraction of the gravitino curvature
\begin{eqnarray}
\label{2.20a}
 \Gamma^b\psi_{ab} & = &
 e^{-1}\phi^3 (\Psi_a +
 {\textstyle{1\over 4}}\sqrt{2}\,\Gamma_a\Lambda)
 - 2\sqrt{2}\,{\cal D}_a\lambda \,.
\end{eqnarray}
 From (\ref{2.20a}) we obtain
 by contracting with a further $\Gamma$-matrix:
\begin{eqnarray}
\label{2.20}
 \Gamma^{ab}\psi_{ab} & = & e^{-1}\phi^3\,(2\Gamma^a\Psi_a
 +{\textstyle{9\over 2}}\sqrt{2}\,\Lambda)
 \,.
\end{eqnarray}

Combining (\ref{2.18}) and (\ref{2.10a})
 one derives two additional identities
 involving the derivative of the gravitino field equation:
\begin{eqnarray}
\label{2.21}
 \fcapslashcD \psi_{ab}
 & = &-2\,{\cal D}_{[a}(e^{-1}\phi^3\Psi_{b]})
 +{\textstyle{1\over 2}}\sqrt{2}\,
 \Gamma_{[a}{\cal D}_{b]}(e^{-1}\phi^3\Lambda)
 \nonumber \\
 & &-{\textstyle{1\over 4}}\Gamma^c\Gamma^{ef}\psi_c\,
 R_{abef}
 -{\textstyle{1\over 2}}\sqrt{2}\,
 \Gamma^{cd}\lambda R_{abcd}
 +\Gamma^c\psi_{[a}R_{b]c} \,, \\
\label{2.22}
 {\cal D}^b\psi_{ab} & = &
 \fcapslashcD (e^{-1}\phi^3\Psi_a)
 -{\textstyle{1\over 4}}\sqrt{2}\,
 \bigl\{\Gamma_a\fcapslashcD (e^{-1}\phi^3\Lambda)
 -{\cal D}_a(e^{-1}\phi^3\Lambda)\bigr\} \nonumber \\
 & &-{\textstyle{1\over 4}}\Gamma^{ef}\psi^b\,R_{abef}
 +{\textstyle{1\over 2}}R_{ab}(\Gamma^{bc}\psi_c-\psi^b)
 +{\textstyle{1\over 4}}R\,\psi_a \,,
\end{eqnarray}
while (\ref{2.17}) and (\ref{2.20}) give:
\begin{eqnarray}
\label{2.17a}
 \fcapslashcD \lambda & = & -e^{-1}\phi^{3}
 \{ {\textstyle{1\over 4}}\sqrt{2}\,\Gamma^a\Psi_a + \Lambda \}
 \,.
\end{eqnarray}
In the identities
 (\ref{2.10a}-\ref{2.17a}) we have not written contributions
 of $H$ and $\phi^{-1}\partial \phi$. In the next section we
 will discuss why these are neglected in our calculations.
\par
In $d=10$, $N=1$ supergravity the only matter multiplet is
 the Yang-Mills multiplet, which consists of the gauge field
 $A_\mu$, and a Majorana-Weyl spinor $\chi$, both in the
 adjoint representation of an arbitrary gauge group.
 The transformation rules are
\begin{eqnarray}
\label{2.25}
 \delta A_\mu & = & {\textstyle{1\over 2}}\bar\epsilon\,
 \Gamma_\mu \chi \,, \\
\label{2.26}
 \delta \chi & = & -{\textstyle{1\over 4}}
 \Gamma^{\mu\nu}F_{\mu\nu}(A)\,\epsilon
 + \epsilon\,({\rm fermi})^2 \,.
\end{eqnarray}
The coupling of the Yang-Mills multiplet $(A_\mu,\chi)$
 to ten-dimensional supergravity \cite{Cham81,ChapMan83}
 leads to
 a supersymmetric action of the form
 ${\cal L}_R+{\cal L}_{F^2}$.
 This requires the inclusion of the Yang-Mills
 Chern-Simons term in the field strength
 (\ref{2.7}) of the $B$-field \cite{ChapMan83}, and a corresponding
 modification of the $B$ transformation rule.
\par
 The cancellation of anomalies requires
 a further modification of
 $H$ by the corresponding Lorentz Chern-Simons term \cite{GS84}.
 However, this mechanism breaks the local supersymmetry.
 The fact that the transformation rules (\ref{2.12})
 and (\ref{2.13}) of
 $\Omega_{\mu}{}^{ab}_-$ and the gravitino curvature $\psi^{ab}$
 have the same structure as those of
 the Yang-Mills multiplet $(A_\mu, \chi)$ (\ref{2.25}), (\ref{2.26})
 simplifies the
 restoration of supersymmetry \cite{BedR89a}.
 By replacing in the action
 $R + \beta \,{\rm tr}\, F^2$,
 and in the corresponding transformation rules
 $A_\mu$ by $\Omega_{\mu}{}^{ab}_-$,
 $\chi$ by $\psi^{ab}$, $F_{\mu\nu}(A)$
 by the corresponding curvature
 $R_{\mu\nu}{}^{ab}(\Omega_-)$, and
 the coupling $\beta$ by an {\em a priori} independent
 coupling $\alpha$, the
 $\,{\rm tr}\,F^2$ Yang-Mills action
 can be
 immediately
 extended to a supersymmetric action of the form
 $R+\beta\,{\rm tr}\,F^2+\alpha R^2$.
 This requires a
 modification, proportional to $\alpha$,
 of the supersymmetry
 transformation rule of the
 $B$-field. Since
 $(\Omega_{\mu}{}^{ab}_-, \psi^{ab})$
 depend on $B$ (see relations (\ref{2.6}) and
 (\ref{2.10})),
 the transformation rules of $\Omega_{\mu}{}_-^{ab}$
 and $\psi^{ab}$ obtain
 order $\alpha$ terms, besides the
 order $\beta$ terms already present due to the
 Yang-Mills coupling. This breaks the invariance
 of the $R+\beta\,{\rm tr}\,F^2+\alpha R^2$-action
 by terms which are of order $\alpha^2$ and
 $\alpha\beta$.
 The best one can hope for in this explicit
 supersymmetrization of the Lorentz
 Chern-Simons term is an iterative invariance
 in the couplings $\alpha$ and $\beta$.
\par
The iterative procedure outlined above was worked out
 for the cubic
 $\alpha^2 R^3$, $\alpha\beta R\,{\rm tr}\,F^2$, and for the
 quartic $\alpha^3R^4$, $\alpha^2\beta R^2\,{\rm tr}\,F^2$,
 $\alpha\beta^2(\,{\rm tr}\,F^2)^2$
 contributions to the supersymmetric effective action.
 Bosonic cubic terms in the supersymmetric action
 are not required.
 Contributions from the variation
 of the quadratic and cubic action play a crucial role
 in the cancellations which lead to the final form of
 the quartic action \cite{BedR89}. Thus the quartic action
 obtained in \cite{BedR89} is directly linked to the
 inclusion of the Lorentz Chern-Simons form, and
 {\em a priori} unrelated to the quartic actions which we will
 construct in this paper, which do not include
 quadratic or cubic contributions.
\newpage
\section{Introduction}

In recent years much work has been devoted to the study of the
 low-energy effective action of string theory. In the limit of low
 energy, string theory can be approximated by ordinary field
 theory, in which string effects should appear as higher derivative
 interaction terms. This effective action provides a useful tool to
 investigate the impact of string theory on particle physics.
\par
In this context, the heterotic string \cite{GrHaMaRo85} is of
 particular interest. Its zero slope limit (the limit in which the
 inverse string tension, $\alpha^\prime$, goes to zero) is given by
 ten-dimensional supergravity coupled to
 Yang-Mills \cite{Cham81,ChapMan83}. Corrections to this
 zero slope limit, proportional to $\alpha^\prime$, are required
 in $d=10,\ N=1$ supergravity to achieve the cancellation of
 anomalies \cite{GS84}. These corrections involve the introduction
 of the Lorentz Chern-Simons term, on the same footing as the
 Yang-Mills Chern-Simons term required by supersymmetry in the
 Einstein-Yang-Mills supergravity theory \cite{ChapMan83}.
\par
One method of investigating the implications of string theory
 for particle physics
 involves the compactification of the effective field theory from
 ten to four dimensions \cite{CaHoStWi85}. The inclusion of the
 Lorentz Chern-Simons term makes it possible to obtain in this way
 phenomenologically interesting models in four
 dimensions\cite{Review}. Supersymmetry in four dimensions, a
 remnant of the space-time supersymmetry of the heterotic string,
 is a common feature of most of these models.
\par
 Much is known about the bosonic contributions
 to the ten-dimensional string
 effective action, ${\cal L}_{\rm eff}$. In this paper we
 investigate the supersymmetric completion of ${\cal L}_{\rm eff}$.
 We may characterize the different contributions to ${\cal L}_{\rm
 eff}$ by the power of the Riemann tensor in the $R^n$-terms which
 they contain:
\begin{eqnarray}
\label{1.1}
 {\cal L}_{\rm eff} & = & \sum_n {\cal L}_{R^n} \,.
\end{eqnarray}
The main issue in this paper is the supersymmetrization of
 the $R^4$-terms in ${\cal L}_{\rm eff}$. Partial results about
 this work were presented in \cite{dRSuWi92}.
\par
Before discussing our results it is useful to present schematically
 what is known about the bosonic part of ${\cal L}_{\rm eff}$. We
 use the results obtained by string amplitude methods. Here one
 calculates string $S$-matrix elements for scattering of massless
 particles, and then reconstructs a field theoretical action which
 reproduces these amplitudes. There are contributions from the
 tree-level (classical) string theory, from one-loop string
 effects, etc. This action is expressed in terms of the physical
 fields of $d=10,\ N=1$ supergravity. The bosonic fields are the
 ten-bein field $e_\mu{}^a$, an antisymmetric tensor gauge
 field $B_{\mu\nu}$ (with field strength $H_{\mu\nu\lambda}$), the
 dilaton field $\phi$, and the Yang-Mills gauge field $A_\mu$ (the
 fermions are introduced in Section 2, where we present some basic
 properties of ten-dimensional supergravity). The presence of the
 dilaton in this action is limited by global scale invariance
 \cite{EWit85}. Our
 fields (except the dilaton) are scale-invariant, while $\phi$
 transforms as $\phi\to \phi\xi$, $\xi$ being the parameter of
 scale transformations. Scale invariance implies that $\phi$ occurs
 only in the combination $\phi^{-1}\partial\phi$, or as an overall
 multiplicative factor in the Lagrangian.
\par
{}From the tree-level string calculation
 \cite{N85,KMN86,GrWi86,CN87,GrSl87}
 one obtains ${\cal L}_R$:
\begin{eqnarray}
\label{LRtree}
 {\cal L}_R &\sim & {1\over \kappa^2}\phi^{-3}\{R + H^2 +
 (\phi^{-1}\partial\phi)^2\},
\end{eqnarray}
where $\kappa$ is the ten-dimensional gravitational coupling
 constant, of dimension $[{\rm mass}]^{-4}$. Also from the string
 tree-level one obtains a quadratic action\footnote{Here $\beta =
 1/(g_{10})^2$, $g_{10}$ the Yang-Mills coupling constant. The
 dimension of $\alpha'$ is $[{\rm mass}]^{-2}$, of $\beta$ $[{\rm
 mass}]^{6}$. The number of string loops is counted by the
 dimensionless coupling $g^2$, which satisfies, for the heterotic
 string, the relation $g = 2\kappa (2\alpha')^{-2}$. $\beta$ is
 fixed by $\beta = \alpha'/(2\kappa^2)$ \cite{GrHaMaRo85}.}:
\begin{eqnarray}
\label{LR2tree}
 {\cal L}_{R^2} &\sim& \phi^{-3}
 \{{\alpha'\over \kappa^2}R^2 + \beta\,{\rm tr}\,F^2 \}\,.
\end{eqnarray}
 and a quartic action\footnote{The absence of the cubic
 action ${\cal L}_{R^3}$ is understood from the vanishing of
 three-point string scattering amplitudes.}
\begin{eqnarray}
\label{LR4tree}
 {\cal L}_{R^4} & \sim & \alpha'\kappa^2\phi^{-3}
 ({\alpha'\over \kappa^2}R^2 + \beta\,{\rm tr}\,F^2)^2
 + {\alpha'{}^3\over \kappa^2}\phi^{-3}\zeta(3) X\,,
\end{eqnarray}
where $X$ is the term \cite{GrvdVZan86a,GrWi86}:
\begin{eqnarray}
\label{LX}
 X = t^{\mu_1\ldots \mu_8}t^{\nu_1\ldots\nu_8}
 R_{\mu_1\mu_2\nu_1\nu_2}
 R_{\mu_3\mu_4\nu_3\nu_4}
 R_{\mu_5\mu_6\nu_5\nu_6}
 R_{\mu_7\mu_8\nu_7\nu_8} \,.
\end{eqnarray}
The tensor $t$ is discussed in Section 3. The transcendental
 coefficient $\zeta(3)$ makes it impossible to relate the two
 contributions in ${\cal L}_{R^4}$ by supersymmetry.
\par
At the one-loop level \cite{SaTa87,ElJeMiz87,AbKuSa88a}
 ${\cal L}_{\rm eff}$ obtains corrections to the
 quartic action:
\begin{eqnarray}
\label{LR4one}
 {\cal L}_{R^4} & \sim &
 \alpha'\kappa^2 g^2\{
 ({\alpha'\over \kappa^2}R^2 + \beta\,{\rm tr}\,F^2)^2
 +\beta^2 \,{\rm tr}\,F^4\}
 + {\alpha'{}^3g^2\over \kappa^2} X\,.
\end{eqnarray}
Note the absence of the factor $\phi^{-3}$ in the one-loop
 contributions. In fact, each string loop will give a
 factor $\phi^{3}g^2$. This can be understood in terms of a
 background field sigma-model calculation from the coupling of the
 dilaton to the Euler character of the world
 sheet \cite{FrTs85,DiSe85,CFMP85}.
\par
Besides the above terms due to four-point scattering amplitudes
 there are also contributions from one-loop five-point
 amplitudes \cite{LeNiSc87,ElMiz88}. These are of the form
\begin{eqnarray}
\label{fivepnt}
 {\cal L}_{R^4} & \sim &
 \epsilon^{\mu_1\ldots\mu_{10}} B_{\mu_1\mu_2}
 \,{\rm tr}\,F_{\mu_3\mu_4}\ldots F_{\mu_9\mu_{10}}\,,
\end{eqnarray}
 while similar terms with $F$ replaced by $R$ also appear.
\par
Other information about the quartic action comes from the
 counterterms
 in the $d=10$ action which are required for anomaly
 cancellations \cite{GS84}.
 We would expect these terms to be part of the string
 effective action. Indeed, terms of the form (\ref{fivepnt})
 are among the counterterms of \cite{GS84}.
 It is then of
 interest to see, whether or not they are linked by supersymmetry
 to some of the terms already present in (\ref{LR4tree})
 and (\ref{LR4one}).
\par
Let us now discuss the supersymmetrization of the effective action.
 The action ${\cal L}_R$ corresponds to the supersymmetric Einstein
 action of $d=10,\ N=1$ supergravity \cite{Cham81}. The inclusion
 of the term $\beta\,{\rm tr}\,F^2$ leads to the supersymmetric
 action of \cite{ChapMan83}. The field strength $H$ then has to be
 extended with the Yang-Mills Chern-Simons term. The introduction
 of the Lorentz Chern-Simons term requires, by supersymmetry, the
 presence of the $R^2$-action. The supersymmetrization of
 the $R^2$-action has been achieved by the Noether
 method \cite{Noether,BedR89a,BedR89} and by
 superspace methods\footnote{For a recent review of superspace
 methods in connection with the Lorentz Chern-Simons terms,
 see \cite{Fre}.} \cite{Italy,Gates}. In \cite{BedR89} an
 explicit supersymmetric action for the Lorentz Chern-Simons term,
 including terms quartic in $R$, was presented. In the absence of
 Yang-Mills couplings it is of the schematic form:
\begin{eqnarray}
\label{LCS}
 {\cal L}_{\rm LCS} & = &
 {\cal L}_R + \phi^{-3}\alpha R^2 + \phi^{-3}\alpha^3 R^4 + \dots
 \,.
\end{eqnarray}
Each term has the same power of $\phi$, and,
 consistent with string amplitude results, the $n=3$ contribution
 is absent. Supersymmetry holds only iteratively in $\alpha$, so
 that the supersymmetry transformation rules of a generic field $V$
 are
\begin{eqnarray}
\label{LCStrans}
 \delta V & = & \sum_{n=0} \alpha^n \delta_n V \,.
\end{eqnarray}
Here $\delta_0 V$ are the transformation rules corresponding to the
 action ${\cal L}_R$. This can easily be generalized to the case
 where Yang-Mills couplings are present. Again schematically, one
 should make everywhere the replacement $\alpha R^2 \to \alpha
 R^2 + \beta \,{\rm tr}\,F^2$. On identifying the {\em a priori}
 independent coupling $\alpha$ with $\alpha'/\kappa^2$ one then
 obtains exactly the terms in the tree-level string amplitude
 result (\ref{LR2tree},\ \ref{LR4tree}), except for
 the $\zeta(3)X$-term.
\par
In this paper we address the problem of supersymmetrizing terms
 quartic in the Riemann tensor. These include the remaining
 tree-level term $\zeta(3)\phi^{-3}X$ and the one-loop
 contributions (\ref{LR4one}). Since the supersymmetrization of
 the $R^2$-terms in ${\cal L}_{\rm eff}$ is complete, this
 supersymmetric $R^4$-action should be of the form
\begin{eqnarray}
\label{Lus}
 {\cal L} & = & {\cal L}_R + \gamma\,R^4 + \ldots \,,
\end{eqnarray}
with modifications to the supersymmetry transformation rules
 of \cite{Cham81,ChapMan83} proportional to $\gamma$.
 Here $\gamma$ is an additional parameter, of dimension
 $[{\rm mass}]^2$, {\em a priori} independent of $\alpha$
 and $\beta$. Relations between $\alpha$, $\beta$ and $\gamma$
 will be required if quartic contributions to ${\cal L}$
 and the string effective action ${\cal L}_{\rm eff}$ are
 to be identified, or if the cancellation of anomalies
 is imposed. Supersymmetry by itself will not relate
 $\alpha$, $\beta$ and $\gamma$.
\par
An obvious problem is already evident from the schematic form of
 the action given above. There are two contributions
 proportional to $X$, one with and one without the
 dilaton-dependent factor. One would expect that supersymmetry
 gives a unique value for the power of $\phi$ which appears in
 front of $X$. The same problem arises for the terms
 with $\alpha R^2 + \beta \,{\rm tr}\,F^2$. In that case one
 should realize however
 that in the tree-level quartic action
 this term is determined
 largely by the presence of $R^2$ in (\ref{LCS}), so that the
 tree-level and one-loop contributions to $(\alpha
 R^2 + \beta \,{\rm tr}\,F^2)^2$ do not appear on the same footing.
\par
A second indication that factors of $\phi$ are important can be
 seen from (\ref{fivepnt}). This term is invariant under gauge
 transformations of the $B$-field only if the factor $\phi^{-3}$ is
 absent. Therefore the presence of the parity-violating
 terms (\ref{fivepnt}) requires the absence of the
 factor $\phi^{-3}$.
\par
As we shall show in this paper the supersymmetrization of {\em any}
 action of the form (\ref{Lus}) requires $\epsilon
 B R^4$ terms, and therefore the absence of $\phi^{-3}$. Thus we
 achieve the supersymmetrization of the one-loop
 contributions (\ref{LR4one}), but not that of the $\zeta(3)$-term
 in (\ref{LR2tree}).
\par
Some results about the supersymmetrization of $R^4$-actions have
 been obtained in superspace \cite{NiTo86,Ka87,LePa89}.
 However, the supersymmetrization of $X$ (\ref{LX})
 in \cite{NiTo86} and \cite{Ka87} depends on an off-shell
 formulation of $d=10, N=1$ supergravity, which has not yet been
 proven to exist. Also, it has not been worked out whether the
 proposed superspace invariant for $X$ represents the tree level
 contribution (\ref{LR4tree}) or the one-loop term
 in (\ref{LR4one}). On the basis of our work we would have to
 conclude that this can only be the one-loop term. Since
 other $R^4$-terms besides $X$ appear in ${\cal L}_{\rm eff}$, we
 prefer to search systematically for the most general
 supersymmetric invariant with the generic structure (\ref{Lus}).
\par
In this paper we use the component field Noether method. One
 starts with an Ansatz for the supersymmetric action
 that one wants to construct. The Ansatz should contain all
 possible terms, each
 with an unknown coefficient. Invariance under supersymmetry is
 then used to determine these coefficients. This method has the
 disadvantage of being algebraically complex. The Ansatz contains
 many terms, so working out the variations involves a large amount
 of work. However, this tedious task can and has all been done by a
 computer program for algebraic manipulations. Then the explicit
 nature of this method turns into an advantage. The resulting
 invariant can be compared in detail with the results from string
 amplitude calculations. Also, the explicit form of the modified
 transformation rules is obtained. The transformation rules of the
 fermions play a crucial role in the study of compactification to
 four dimensions \cite{CaHoStWi85}.
\par
The full calculation will be done for the gravitational sector
 only, i.e., without the Yang-Mills coupling. We shall see that our
 results can be generalized
 to the case were the
 Yang-Mills multiplet is present as well.
\par
This paper is organized as follows. In Section 2 we present some
 basic material on $d=10,\ N=1$ supergravity. We also briefly
 discuss results about the supersymmetric $R^2$-action. In Section 3
 we construct the Ansatz (given explicitly in Appendix A) for the
 supersymmetric $R^4$-action. Of course, for practical reasons we
 have to limit ourselves to certain sectors of the complete
 action (for instance, we never include four-fermion terms). These
 limitations are also discussed in Section 3. In Section 4 we give
 a schematic overview of the calculation, and consider in some
 detail a particular sub-calculation which leads us to conclude
 that terms such as (\ref{fivepnt}) must be present in the final
 result.
 The full result, and its generalization to the Yang-mills case, is
 then presented in Section 5 and Appendix B. Section 6 compares our
 results with the string amplitude calculations and discusses the
 relation with other work.

\newpage
\section{$R^4$--Invariants and the Ansatz}
\vskip 12pt
The supersymmetrization of $R^4$-action starts with the
 construction of an Ansatz, which should contain all terms that
 might be linked to the $R^4$-terms by supersymmetry.
\par
In order to make the supersymmetrization feasible one has to
 put restrictions on the terms which are included
 in the Ansatz, and, correspondingly, on the contributions
 to its supersymmetry variation. In this section we will
 discuss the structure of our Ansatz and the restrictions
 we have imposed.
\par
As we have already mentioned in Section 2, we will not consider
 terms in the action which are quartic in fermions.
 Hence, in the $R^4$-action only purely bosonic terms
 and terms quadratic in fermions will appear. Correspondingly,
 in the supersymmetry transformations of the bosonic fields only
 terms
 linear in fermions, in the transformations of
 the fermionic fields only the bosonic
 contributions have to be considered:
\begin{eqnarray}
 \delta({\rm boson}) & = &
 \bar\epsilon\,({\rm fermion})\,,
 \nonumber\\
 \delta({\rm fermion}) & = &
 ({\rm boson})\,\epsilon\,. \nonumber
\end{eqnarray}
\par
In the $R^4$-action we do not write terms which
 contain the equations of motion of the $R$-action
 (\ref{2.15}--\ref{2.19}).
 Such contributions
 can always be eliminated
 by a suitable redefinition of the
 corresponding field.
 As was outlined in \cite{GrWi86,GrSl87}, the
 results obtained from scattering amplitude calculations
 are insensitive to such redefinitions of the
 fields. Thus, we do not have to
 include terms in the Ansatz containing a Ricci tensor or
 a curvature scalar.
 The same applies to terms
 containing a contracted derivative of the Riemann tensor, since
\begin{eqnarray}
\label{3.0}
 {\cal D}_\mu(\omega) R_{\lambda\rho}{}^{\mu a}(\omega)
 & = & 2\, {\cal D}_{[\lambda}(\omega) R_{\rho]}{}^a(\omega) \,.
\end{eqnarray}
 Similarly, fermionic terms
 containing the left-hand-side of
 (\ref{2.20a}-\ref{2.17a}) can be left out.
\par
The presence of the fields $\phi$ and $B_{\mu\nu}$ in $d=10$
 supergravity complicates our calculations considerably. The
 occurrence of $B_{\mu\nu}$ itself is of course restricted by
 the requirement of gauge invariance (see (\ref{2.8})),
 but many contributions containing the field strength $H$ are
 possible.
 One may attempt to restrict the contributions of $H$ by
 requiring that $H$ only occurs as torsion (\ref{2.6}), as
 seems to be indicated by string amplitude calculations.
 However, we prefer not to bias our calculations by introducing
 such input.
 Similarly,
 the appearance of $\phi$ can be restricted by requiring
 global scale invariance, but $\phi^{-1}\partial\phi$ may
 appear anywhere.
\par
We compromise by including in the action only terms
 independent of or linear in $H$
 and $\phi^{-1}\partial\phi$. In the variation of the action
 we should then consider only those terms in which $H$ and
 $\phi^{-1}\partial\phi$ are absent.
 From (\ref{2.4})
 we see that this implies for instance that we never have
 to vary the field $\lambda$, and that consequently there is no need
 to include $(\lambda)^2$-terms in the action. Furthermore,
 we can restrict the terms containing $H$ and
 $\phi^{-1}\partial\phi$
 to be purely bosonic.
\par
In the Ansatz we use the spin connection with $\psi$-torsion,
 i.e., $\omega_\mu{}^{ab}(e,\psi)$, as the argument of the
 Riemann tensor, and parametrize the terms
 linear in $H$ separately. In the $H$-dependent terms in the
 Ansatz we use the supercovariant field strength $\hat H$,
 given in (\ref{2.7}).
\par
Note that with the above restrictions, it is no longer guaranteed
 that our method will yield a useful result. It may well be, for
 instance,
 that the cancellation of variations containing $H$ are
 required to fix the coefficients of the terms linear in $H$ in
 the action uniquely. As the next sections will show,
 a large part of the supersymmetric action is determined, even
 though we do not consider the cancellation of all possible
 variations.
 More fundamentally, we must
 admit that our method does not strictly prove the existence of
 a supersymmetric invariant, since the procedure may still fail
 for variations which we do not consider. The
 results, and their
 relation with string amplitude calculations,
 give us confidence that our procedure could in principle be
 continued to the end without essential obstructions.
\vskip8pt

The purpose of our present work is the supersymmetrization of
 $R^4$-actions, with in view the application to the
 effective action of heterotic string theory.
 As discussed in the Introduction,
 the bosonic part corresponding to
 tree level and one-loop contributions to
 string amplitudes are known.
 There, the
 following actions quartic in the Riemann tensor arise:
\begin{eqnarray}
\label{3.1}
 X & = & t^{\mu\nu\lambda\rho\sigma\tau\alpha\beta}
 t_{abcdefgh}R_{\mu\nu}{}^{ab}
 R_{\lambda\rho}{}^{cd}
 R_{\sigma\tau}{}^{ef}
 R_{\alpha\beta}{}^{gh} \,, \\
\label{3.2}
 Y_1 & = & t^{\mu\nu\lambda\rho\sigma\tau\alpha\beta}
 R_{\mu\nu}{}^{ab}R_{\lambda\rho}{}^{ab}
 R_{\sigma\tau}{}^{cd}R_{\alpha\beta}{}^{cd} \,, \\
\label{3.3}
 Y_2 & = & t^{\mu\nu\lambda\rho\sigma\tau\alpha\beta}
 R_{\mu\nu}{}^{ab}R_{\lambda\rho}{}^{bc}
 R_{\sigma\tau}{}^{cd}R_{\alpha\beta}{}^{da} \,, \\
 \label{3.4}
 Z & = & R_{[ab}{}^{ab}R_{cd}{}^{cd}R_{ef}{}^{ef}R_{gh]}{}^{gh} \,.
\end{eqnarray}
 The tensor $t$ has the following structure when acting on
 commuting,
 antisymmetric tensors\footnote{In string amplitude considerations
 (see e.g.
 \cite{GrSl87}) the indices of the $t$-tensor indicate the
 eight transverse directions in light-cone coordinates, and then $t$
 contains an additional eight-dimensional
 Levi-Civit\'a symbol.
 Here we extend the range of the indices to all ten
 values.}$M_i$, $i=1,..,4$:
\begin{eqnarray}
\label{3.5}
 t_{abcdefgh}M_1^{ab}M_2^{cd}M_3^{ef}M_4^{gh} & = &
 -2\bigl\{{\rm tr}M_1M_2{\rm tr}M_3M_4
 +{\rm tr}M_2M_3{\rm tr}M_4M_1 \nonumber \\
 & &\phantom{12345678}
 +{\rm tr}M_1M_3{\rm tr}M_2M_4 \bigr\} \nonumber \\
 & & +8\bigl\{{\rm tr}M_1M_2M_3M_4
 +{\rm tr}M_1M_3M_2M_4 \nonumber \\
 & &\phantom{12345678}
 +{\rm tr}M_1M_3M_4M_2\bigr\} \,.
\end{eqnarray}
\par
The action (\ref{3.1}) was obtained
 from a calculation of the two-loop $\beta$-function in a
 supersymmetric nonlinear sigma-model \cite{GrvdVZan86a}
 and independently in string amplitude
 calculations \cite{GrWi86}.
 This action appears
 in the
 tree level string effective action with a
 characteristic coefficient $\zeta(3)$.

The action $Y_1$, which has the structure
 $t^{\ldots}({\rm tr}R^2)^2$, was also found
 in tree-level string amplitude calculations\footnote{For
 comparison to tree-level string amplitude
 results we will use the very detailed result
 given in \cite{GrSl87}.} \cite{GrSl87}. Note
 that $Y_2$ has a different trace structure
 $t^{\ldots}({\rm tr}R^4)$. Finally,
 (\ref{3.4}) is invariant under linearized supersymmetry
 transformations, since by the Bianchi identity of the Riemann
 tensor,
\begin{eqnarray}
\label{Bianchi}
 {\cal D}_{[\mu}(\omega)R_{\nu\rho]}{}^{ab}(\omega) & = & 0 \,,
\end{eqnarray}
 the variation of $Z$ is a total derivative for any variation of
 $\omega$.
 If $Z$ is reduced to eight dimensions
 it becomes a total derivative.
 This implies that it does not
 play a role in lightcone gauge string amplitude calculations.
 Therefore one has no {\em a priori} knowledge
 from string amplitude
 or sigma model calculations about its effects in a ten-dimensional
 supersymmetric invariant. The fact that in ten dimensions one
 should allow the inclusion of a $Z$-action was emphasized in
 \cite{Zu86,NiGa87,My87}.
\par
In the supersymmetrization of $R^4$-actions we look for
 invariants of the form
\begin{eqnarray}
\label{gamma}
 {\cal L} & = & R+\gamma R^4+O(\gamma^2)\,,
\end{eqnarray}
where $ R$ is the pure $d=10$, $N=1$ supergravity action
 (\ref{2.14}).
Supersymmetry may hold iteratively in $\gamma$, so that
 the supergravity fields will need modifications of the
 supersymmetry transformation rules of $O(\gamma)$ in order to
 achieve invariance of the action (\ref{gamma}) to $O(\gamma)$.
\vskip 8pt
Our Ansatz in the search for the supersymmetric completion
 of $R^4$-actions is written in the form:
\begin{eqnarray}
\label{Ltot}
{\cal L}_{\rm tot} & = & \gamma\,e\phi^{y}\sum_i{\cal L}_i \,.
\end{eqnarray}
 The sum is over the different structures that may occur in the
 action.
 We consider 15 different sectors, which are presented in Appendix
 A,
 four involving purely
 bosonic terms (${\cal L}_1$ -- ${\cal L}_4$), four
 sectors involving the gravitino field $\psi_\mu$ and its curvature
 $\psi_{(2)}$ (${\cal L}_5$ -- ${\cal L}_8$)
 and seven sectors containing the dilatino field $\lambda$ (${\cal
 L}_9$ --
 ${\cal L}_{15}$). We give a few comments on our Ansatz.
\par
We include an arbitrary power of the dilaton in front of the
 action (the $R$-action also has such a structure, with
 $\phi^{-3}$).
 Note that by supersymmetry the power of $\phi$ has
 to be independent of the index $i$ labelling the different sectors,
 since the supersymmetry transformation rules (\ref{2.1}--\ref{2.5})
 contain no explicit powers of $\phi$.
\par
The sector ${\cal L}_1$
 (\ref{L1}) contains all possible contractions of four Riemann
 tensors. Therefore, the actions (\ref{3.1}-\ref{3.4}) can be
 written as
 linear combinations of the terms given in (\ref{L1}).
 Using pair exchange and cyclic identities
 for the Riemann tensor,
 and neglecting terms containing the Ricci tensor or
 curvature scalar,
 one finds\footnote{A capital letter denotes the term
 in the Ansatz without the corresponding parameter (always given
 in lower case) (see Appendix A).}:
\begin{eqnarray}
\label{3.6}
 X & = & 12\bigl\{A_1-16A_2+2A_3-32A_5+16A_6+32A_7\bigr\} \,,
 \nonumber \\
 Y_1 & = & -2A_1+16A_2-4A_3+8A_4 \,, \nonumber \\
 Y_2 & = & -4A_2+2A_4-16A_5+8A_6+16A_7 \,, \nonumber \\
 Z & = &
 {\textstyle{1\over 7\times 5!}}
 \{A_1-16A_2+2A_3+16A_4-32A_5+16A_6-32A_7\} \,.
\end{eqnarray}
Note that the actions $X$, $Y_1$ and $Y_2$ are related by
\begin{eqnarray}
\label{XYZ-relation}
 X+6Y_1-24Y_2 & = & 0 \,.
\end{eqnarray}
The sector ${\cal L}_1$ is the only one
 for which the variation of $\phi$ in front
 of the action has to be evaluated.
\par
The ${\cal L}_3$-sector consists of the following two terms:
\begin{eqnarray}
\label{K1}
 K_1 & = & ie^{-1}\epsilon^{\mu_1...\mu_{10}}B_{\mu_1\mu_2}
 R_{\mu_3\mu_4}{}^{ab}R_{\mu_5\mu_6}{}^{ab}
 R_{\mu_7\mu_8}{}^{cd}R_{\mu_9\mu_{10}}{}^{cd} \,, \\
\label{K2}
 K_2 & = & ie^{-1}\epsilon^{\mu_1...\mu_{10}}B_{\mu_1\mu_2}
 R_{\mu_3\mu_4}{}^{ab}R_{\mu_5\mu_6}{}^{ac}
 R_{\mu_7\mu_8}{}^{bd}R_{\mu_9\mu_{10}}{}^{cd} \,.
 \end{eqnarray}
Both terms are clearly invariant under gauge transformations
 (\ref{2.8}) of the
 $B$-field
 because of the Bianchi identity (\ref{Bianchi}).
 Note that this gauge invariance requires the
 absence of the dilaton field
 in (\ref{K1}) and (\ref{K2}), i.e.,
 $y=0$ in (\ref{Ltot}).
 One-loop string amplitude calculations reveal
 that these $K$-terms must be part of the effective
 string action
 \cite{LeNiSc87,ElMiz88}.
\par
The sectors ${\cal L}_5$ -- ${\cal L}_8$ parametrize terms of type
 $\bar\psi_{(2)}\Gamma\psi_{(2)}R{\cal D}R$,
 $\bar\psi_{(2)}\Gamma{\cal
 D}\psi_{(2)}R^2$, $\bar\psi\Gamma\psi_{(2)}R^3$
 and $\bar\psi\Gamma\psi R^2{\cal D}R$ respectively. As we
 noted above, in constructing these sectors we do not allow terms
 with any
 contractions of the form (\ref{2.20a}-\ref{2.22}).
 Note that a partial integration and the use of
 the Bianchi identity (\ref{2.10a}) may relate terms of these
 sectors.
 Therefore, in order to find a minimal set of independent terms for
 the Ansatz
 only those terms are taken into account which are not related by
 any of these
 operations.
\par
Similar arguments apply for the $\lambda$-sectors ${\cal L}_9$ --
 ${\cal
 L}_{15}$. There we do not write terms which are related to the
 equation of motions $\Lambda$ (\ref{2.17}) or $\Psi_\mu$
 (\ref{2.18}).

\newpage
\section{The Calculation}
In this section we will discuss some of the technical aspects
 concerning the calculation we have outlined in the
 previous section.
\par
In Table 1 we present a schematic form of the supersymmetry
 transformations
 relevant for our purposes. Their precise form is given in Section
 2. Note
 that due to the restrictions
 we have imposed we may refrain from considering
 various other contributions such as $\delta\omega =
 \bar\epsilon\,\psi H$.
\begin{table}[h]
\centerline{
\vbox{%
\offinterlineskip
\def\mr{\omit&height 3pt&&&&}
\halign{\strut#&
 \vrule#\quad&
 \hfil#\hfil&
 \quad\vrule#\quad&
 $#$\hfil&
 \quad\vrule#\cr
 \noalign{\hrule} \mr \cr
 &&\# &&\omit \bf Transformation & \cr
 \mr \cr\noalign{\hrule} \mr \cr
 && 1 && \delta\psi = {\cal D}(\omega)\epsilon & \cr
 && 2 && \delta H =\bar\epsilon\,\psi_{(2)},\,
 \delta B = \bar\epsilon\,\psi & \cr
 && 3 && \delta\omega = \bar\epsilon\,\psi_{(2)}
 & \cr
 && 4 && \delta\psi_{(2)} = \epsilon \, R & \cr
 && 5 && \delta e = \bar\epsilon\,\psi
 & \cr
 && 6 && \delta\phi = \bar\epsilon\,\lambda
 & \cr
 && 7 && \delta(\phi^{-1}\partial\phi)=
 \bar\epsilon\,{\cal D}(\omega)\lambda
 & \cr
 \mr\cr\noalign{\hrule}
}}}
\begin{keylistone}
 \item[{\it Table 1.}]{\small The schematic form of the
 supersymmetry
 transformation rules considered in this
 paper. The symbol $\psi$ represents the gravitino,
 $\psi_{(2)}$ the gravitino curvature.}
\end{keylistone}
\end{table}
\par
Table 2 shows the generic structure of the variations of the action
 that
 emerge when applying the transformations (1) -- (7) to the Ansatz.
 In calculating the variation of the Ansatz we always integrate
 away
 from the supersymmetry parameter $\epsilon$ by performing a
 partial
 integration. The variation is then simplified by working out
 products of $\Gamma$-matrices, etc., and brought to a standard
 form. The result then has to vanish, which determines the unknown
 coefficients.
\par
In many cases however, contributions to a variation do not
 have to cancel against each other.
 If a variation is proportional to one of the equations
 of motion (\ref{2.15}--\ref{2.19}) it can be cancelled by changing
 the transformation rule of the corresponding field with a
 contribution of $O(\gamma)$. Consider for example a variation
 which is
 of the form
\begin{eqnarray}
\label{cancel1}
 \delta{\cal L}_{\rm tot} & =& \gamma\, \bar\epsilon\,O^\mu\Psi_\mu
 \,,
\end{eqnarray}
where $O^\mu$ is
a field dependent object which may contain $\Gamma$-matrices.
 Since $\Psi_\mu$ is the
gravitino equation of motion of the action ${\cal L}_R$, a
 variation
$\delta_\gamma\bar\psi_\mu$ of the gravitino in ${\cal L}_R$ with
 parameter
$-\gamma\, \bar\epsilon\,O_\mu$ will give
\begin{eqnarray}
\label{cancel2}
 \delta_{\gamma}{\cal L}_R & = & -\gamma\,
 \bar\epsilon\,O^\mu\Psi_\mu \,.
\end{eqnarray}

\newpage
\centerline{
\vbox{%
\offinterlineskip
\def\mr{\omit&height 3pt&&&&&&&&}
\halign{\strut#&
 \vrule#\quad&
 \hfil#\hfil&
 \quad\vrule#\quad&
 $#$\hfil&
 \quad\vrule#\quad&
 \hfil#\hfil&
 \quad\vrule#\quad&
 #\hfil&
 \quad\vrule#\cr
 \noalign{\hrule} \mr \cr
 &&\#&&\omit Variation && Identity && Cancelled by & \cr
 \mr \cr\noalign{\hrule} \mr \cr
 && (A) &&\bar\epsilon\,\psi_{(2)}R^2{\cal D}R && (\ref{2.20a}),
 (\ref{2.20}) && $\delta\psi$, $\delta\lambda$ & \cr
 && (B) &&\bar\epsilon\,{\cal D}\psi_{(2)} R^3 && (\ref{2.10a}),
 (\ref{2.21})--(\ref{2.22})&& (I), (J), $\delta\psi$,
 $\delta\lambda$ & \cr
 && (C) &&\bar\epsilon\,\psi\, R({\cal D}R)^2
 && -- && -- & \cr
 && (D) &&\bar\epsilon\,\psi\, R^2{\cal D}{\cal D}R && (\ref{2.24})
 &&
 (I) & \cr
 && (E) &&\bar\epsilon\,{\cal D}\lambda\, R^2{\cal D}R &&
 (\ref{2.17a}) &&
 $\delta\lambda$, $\delta\psi$ & \cr
 && (F) &&\bar\epsilon\,{\cal D}{\cal D}\lambda\, R^3 &&
 (\ref{2.17a}),
 (\ref{2.23}) &&
 (J), $\delta\lambda$& \cr
 && (G) &&\bar\epsilon\,\lambda \, R({\cal D}R)^2
 && -- && -- & \cr
 && (H) &&\bar\epsilon\,\lambda \, R^2{\cal D}{\cal D}R &&
 (\ref{2.24}) &&
 (J) & \cr
 && (I) &&\bar\epsilon\,\psi \, R^4
 && -- && -- & \cr
 && (J) &&\bar\epsilon\,\lambda \, R^4
 && -- && -- & \cr
 \mr\cr\noalign{\hrule}
}}}
\begin{keylistone}
\item[{\it Table 2.}]{\small The different structures in the
 variation of
 the action.
 The third column indicates identities used to rewrite
 various contributions. The last column shows how
 these contributions are cancelled. A $\delta\psi$
 or a $\delta\lambda$ entry indicates a
 modification of the transformation rules of the
 corresponding fermion.}
\end{keylistone}
\centerline{
\vbox{%
\offinterlineskip
\def\mr{\omit&height 3pt&&&&&&&&&&&&&&&&&&&&&&}
\halign{\strut#&
 \vrule#\quad&
 \hfil$#$\hfil&
 \quad\vrule#\quad&
 \hfil#\hfil&
 \quad\vrule#\quad&
 \hfil#\hfil&
 \quad\vrule#\quad&
 \hfil#\hfil&
 \quad\vrule#\quad&
 \hfil#\hfil&
 \quad\vrule#\quad&
 \hfil#\hfil&
 \quad\vrule#\quad&
 \hfil#\hfil&
 \quad\vrule#\quad&
 \hfil#\hfil&
 \quad\vrule#\quad&
 \hfil#\hfil&
 \quad\vrule#\quad&
 \hfil#\hfil&
 \quad\vrule#\quad&
 \hfil#&
 \quad\vrule#\cr
 \noalign{\hrule} \mr \cr
 &&{\cal L}_i
 &&(A)&&(B)&&(C)&&(D)&&(E)&&(F)&&(G)&&(H)&&(I)&&(J) & \cr
 \mr \cr\noalign{\hrule} \mr \cr
 &&{\cal L}_1
 && 3 && - && - && - && - && - && - && - && 5 && 6 & \cr
 &&{\cal L}_2
 && 2 && - && - && - && - && - && - && - && - && - & \cr
 &&{\cal L}_3
 && - && - && - && - && - && - && - && - && 2 && - & \cr
 &&{\cal L}_4
 && - && - && - && - && 7 && - && - && - && - && - & \cr
 &&{\cal L}_5
 && 4 && - && - && - && - && - && - && - && - && - & \cr
 &&{\cal L}_6
 && 4 && 4 && - && - && - && - && - && - && - && - & \cr
 &&{\cal L}_7
 && 1 && 1 && - && - && - && - && - && - && 4 && - & \cr
 &&{\cal L}_8
 && 1 && - && 1 && 1 && - && - && - && - && - && - & \cr
 &&{\cal L}_{9}
 && - && - && - && - && - && - && - && - && - && 4 & \cr
 &&{\cal L}_{10}
 && - && - && - && - && 1 && - && 1 && 1 && - && - & \cr
 &&{\cal L}_{11}
 && - && - && - && - && 1 && 1 && - && - && - && - & \cr
 &&{\cal L}_{12}
 && - && - && - && - && - && 4 && - && - && - && - & \cr
 &&{\cal L}_{13}
 && - && - && - && - && 4 && - && - && - && - && - & \cr
 &&{\cal L}_{14}
 && - && - && - && - && - && - && - && 4 && - && - & \cr
 &&{\cal L}_{15}
 && - && - && - && - && - && - && 4 && - && - && - & \cr
 \mr\cr\noalign{\hrule}
}}}
\begin{keylistone}
\item[{\it Table 3.}] {\small All contributions to the variations
 considered
 in Table 2. The numbers in the table
 correspond to the
 supersymmetry transformations given in Table 1. The
 ${\cal L}_i$-entries denote the different sectors
 of the Ansatz, given in the Appendix.}

\end{keylistone}
\newpage
\noindent This new
transformation rule of the gravitino cancels (\ref{cancel1})
in the variation of ${\cal L}_R + {\cal L}_{{\rm tot}}$.
The new transformation applied to ${\cal L}_{\rm tot}$ gives a
contribution proportional to $\gamma^2$, which we need not
consider in this stage of our procedure.
\par
If a variation of ${\cal L}_{\rm tot}$ can be rewritten using
 an identity
 such as (\ref{2.10a}) then its contribution is
 shifted to another part of the calculation.
 Besides (\ref{2.10a}) one
 also has the useful relations
\begin{eqnarray}
\label{2.23}
 {\cal D}_{[\mu}{\cal D}_{\nu]}\lambda
 & = & -{\textstyle{1\over 8}}\Gamma^{ab}\lambda
 R_{\mu\nu ab} \,, \\
\label{2.24}
 {\cal D}_{[\mu}{\cal D}_{\nu]}R_{abcd}
 & = & R_{\mu\nu[a}{}^fR_{b]fcd}
 -R_{\mu\nu[\underline{c}}{}^fR_{abf\underline{d}]}
 \,.
\end{eqnarray}
In some cases, using identities such
 as (\ref{2.20a}--\ref{2.22}), a contribution can be rewritten in
 terms of equations of motion and additional terms which contribute
 to other variations. This mechanism is indicated in the third
 and fourth columns of Table 2. In the fourth column we have not
 indicated explicitly cancellation through modifications to the
 transformation rule of the tenbein. Ricci tensors occur,
 either directly
 or through (\ref{3.0}), in all the variations (A)--(J).
\par
The basic tactic is then to shift as much as possible of a
 particular variation to equations of motion and/or the
 variations (I) and (J) of Table 2,
 by using the identities
 mentioned in the third column of the table. Everything which
 cannot be shifted, which is true in particular for all
 contributions to the variations (I) and (J), has to cancel
 and is used to fix coefficients.
 Table 3 indicates how the different sectors of the Ansatz
 contribute,
 through the supersymmetry transformations of Table 1, to the
 variations of Table 2.
\par
As an example consider
 variations of type (B), i.e. $\bar\epsilon\,{\cal
 D}\psi_{(2)}R^3$.
 From Table 3 we see that these variations are generated by the
 sectors
 ${\cal L}_6$ and ${\cal L}_7$. From ${\cal L}_6$, the
 $\bar\psi_{(2)}\Gamma{\cal D}\psi_{(2)}R^2$-terms, we obtain (B)
 by varying the first gravitino curvature, which is the
 transformation
 numbered 4 in Table 1. From ${\cal L}_7$, the Noether terms
 $\bar\psi\Gamma\psi_{2}R^3$, we find this variation by varying
 the gravitino, and taking, after the partial integration away
 from $\epsilon$, the contribution containing
 ${\cal D}\psi_{(2)}$. This is transformation 1. On simplifying
 these
 variations we isolate those contributions which can be
 written as a Bianchi identity (\ref{2.10a}), or which take
 on the form (\ref{2.21}-\ref{2.22}). This gives
 variations of type (I) and (J) ((J) only in the case (\ref{2.21})
 is used) and equations of motion. Note that in the variations (B)
 we will not encounter the left-hand-side
 of (\ref{2.20a}-\ref{2.20}).
 Such contractions between $\psi_{(2)}$ and $\Gamma$-matrices are
 absent in the Ansatz, as explained in section 3. Contributions
 containing (\ref{2.20a}-\ref{2.20})
 would therefore have to come from the variation of the ${\cal
 L}_6$ terms,
 but it is easy to see that the products of $\Gamma$-matrices
 in these variations do
 not involve the indices of the gravitino curvature.
\par
Of course also the bosonic equations of motion, and the Bianchi
 identity
 (\ref{Bianchi}) are used in the same way. However, the use of these
 does not generate a remainder.
\par
An important role in the calculation is played by the two $K$-terms
 (\ref{L3}).
 If they are part of the action, the power of the dilaton in front
 of
 the action (\ref{Ltot}) will have to vanish. We find that indeed
 the presence of the $K$-terms is unavoidable. Interestingly,
 this result can be seen relatively easily, since only a few terms
 in the Ansatz interact with the $K$-terms. As an example, which
 also
 illustrates explicitly our procedure\footnote{Except for the
 fact that the algebraic manipulations in the following calculation
 were of course performed by our computer program!},
 we will work out the
 contribution of the $K_1$-term.
\par
In the variation of $K_1$ we only have to consider the
 transformation
 of the field $B$, (\ref{2.3}). The $\epsilon$-tensor and the
 $\Gamma$-matrix are combined to give:
\begin{eqnarray}
\label{deltaK1}
 \delta K_1 & = &
 - {\textstyle{1\over 2}}\sqrt{2}\,
 R_{mn}{}^{ab}R_{pqab} R_{rs}{}^{cd}R_{tucd}
 \bar\epsilon\,\Gamma^{mnpqrstuv}\psi_v \,.
\end{eqnarray}
The only term in the Ansatz which gives rise to a similar variation
 is
 $M_{106}$ in (\ref{L7}). In $M_{106}$ we have to vary the gravitino
 and the gravitino curvature. After a partial integration, and
 upon using the Bianchi-identity (\ref{Bianchi}) we find
\begin{eqnarray}
\label{deltaM106}
 \delta M_{106} & = &
 - R_{mn}{}^{ab}R_{pqab} R_{rs}{}^{cd}
 \bar\epsilon\,\Gamma^{mnpqrst}{\cal D}_t \psi_{cd} \nonumber\\
 && +{\textstyle{1\over 4}}
 R_{mn}{}^{ab}R_{pqab} R_{rs}{}^{cd}R_{tucd}
 \bar\epsilon\,\Gamma^{tu}\Gamma^{mnpqrsv}\psi_v \,.
\end{eqnarray}
In the first term we extract $\Gamma^t$ from the $\Gamma$-matrix,
 using
\begin{eqnarray}
 \Gamma^{mnpqrst} = \Gamma^{mnpqrs}\Gamma^{t}
 - 6 \Gamma^{[mnpqr} \delta^{s]t}\,. \nonumber
\end{eqnarray}
Thus we obtain
\begin{eqnarray}
 - R_{mn}{}^{ab}R_{pqab} R_{rs}{}^{cd}
 \bar\epsilon\,\Gamma^{mnpqrs}\fcapslashcD \psi_{cd} \,,\nonumber
\end{eqnarray}
and other terms, which will never contribute to a variation with
 a nine-index $\Gamma$-matrix. Now we use (\ref{2.21}) to
 obtain terms proportional to equations of motion, as well as
\begin{eqnarray}
\label{M106var1}
 +{\textstyle{1\over 4}} R_{mn}{}^{ab}R_{pqab} R_{rs}{}^{cd}R_{tucd}
 \bar\epsilon\,\Gamma^{mnpqrs}
 \Gamma^{v}\Gamma^{tu}\psi_{v} \,.
\end{eqnarray}
We now work out all the products of $\Gamma$-matrices in the
 second term in
 (\ref{deltaM106}) and (\ref{M106var1}),
 and finally obtain the following
 contribution with a $\Gamma^{(9)}$:
\begin{eqnarray}
\label{M106var}
 \delta M_{106} & = &
 {\textstyle{1\over 2}}
 R_{mn}{}^{ab}R_{pqab} R_{rs}{}^{cd}R_{tucd}
 \bar\epsilon\,\Gamma^{mnpqrstuv}\psi_v \,.
\end{eqnarray}
The contributions (\ref{deltaK1}) and (\ref{M106var}) must cancel,
 since
 none of the other terms in the Ansatz produces such a variation.
 Therefore
\begin{eqnarray}
\label{k1eq}
 k_1 & = & {\textstyle{1\over 2}}\sqrt{2}\,m_{106}\,.
\end{eqnarray}
To find out whether or not a term of type $K_1$ is present we
 therefore have to know the value of $m_{106}$. This coefficient
 is determined by considering the following two variations:
\begin{eqnarray}
\label{G3var}
 R_{mn}{}^{ab}R_{pqab}R_{stcd}R^{stcd}\,
 \bar\epsilon\,\Gamma^{mnpqr}\psi_r \,, \\
\label{G1var}
 R_{mnab}R^{mnab}R_{stcd}R^{stcd}\,
 \bar\epsilon\,\Gamma^{r}\psi_r \,.
\end{eqnarray}
To (\ref{G3var}) we get contributions from $M_{106}$, on working out
 the product of $\Gamma$-matrices in (\ref{deltaM106}) and
 (\ref{M106var1}). We also get contributions from $M_{30}$. By
 a calculation similar to the one outlined above for $M_{106}$,
 using the equation of motion (\ref{2.21}),
 we get two equal contributions from $M_{30}$. We then find
 $m_{30} = 2m_{106}$. Finally we calculate the contributions
 to the variation (\ref{G1var}). These come from the previous
 calculation of the variation of $M_{30}$, and also from the
 tenbein variation in $A_1$. The result is that $m_{30} =
 {\textstyle{1\over 2}} a_1$. Thus we conclude that this
 calculation determines $k_1$:
\begin{eqnarray}
\label{k1val}
 k_1 & = & {\textstyle{1\over 8}}\sqrt{2}\, a_1 \,.
\end{eqnarray}
The presence of the $K_1$-term is therefore linked
 by supersymmetry to the
 presence of $A_1$. The possibility of having $a_1=0$ will
 be discussed in the next section.
 None of the other terms in the Ansatz contributes to
 (\ref{G3var}) or (\ref{G1var}). The feature which singles
 out these variations is the contraction between the index
 of the gravitino and the $\Gamma$-matrix. Such a contraction
 can only arise from the variation of the tenbein in the $A$-terms
 (\ref{L1}),
 or from terms in ${\cal L}_7$ (\ref{L7}), which already have such a
 contraction. A glance at such terms in the Ansatz shows
 that indeed only $M_{106}$ and $M_{30}$ have the appropriate
 structure.
\par
The $K_1$-term is only invariant under the gauge transformations
 of the $B$-field, if the factor dependent on the
 dilaton in (\ref{Ltot}) is absent. We expect then,
 given the presence of the $K$-term, that supersymmetry will fix
 $y=0$. To see this we will consider variations of type
 (J), $\bar\epsilon\,\lambda R^4$. There are three variations which
 play a
 determining role in fixing the value of $y$. These are
\begin{eqnarray}
\label{Gam0}
 R_{mnab}R^{mnab}R_{stcd}R^{stcd}\,
 \bar\epsilon\,\lambda \,, \\
\label{Gam4}
 R_{mn}{}^{ab}R_{pqab}R_{stcd}R^{stcd}\,
 \bar\epsilon\,\Gamma^{mnpq}\lambda \,, \\
\label{Gam6}
 R_{mn}{}^{ab}R_{pqab}R_{rscd}R^{tucd}\,
 \bar\epsilon\,\Gamma^{mnpqrstu}\lambda \,.
\end{eqnarray}
To these variations we will get contributions from $M_{106}$ and
 $M_{30}$.
 These arise from the use of (\ref{2.21}) in (\ref{deltaM106}) and
 in the related variation of $M_{30}$. Then there are contributions
 from (\ref{L9}), in particular from $P_1$ and $P_{21}$, obtained
 from the variation of the gravitino curvature $\psi_{(2)}$.
 Finally, there
 is of course a contribution to (\ref{Gam0}) from the variation
 of $\phi^y$ in front of the $A_1$-term. The resulting equations
 for the coefficients read
\begin{eqnarray}
 ({\rm \ref{Gam0}}) & \rightarrow &
 -{\textstyle{1\over 3}} y \sqrt{2}\, a_1
 - \sqrt{2}\, m_{30}
 + {\textstyle{1\over 2}}p_1 = 0 \,, \nonumber \\
 ({\rm \ref{Gam4}}) &\rightarrow &
 + 2\sqrt{2}\, m_{30} - 4\sqrt{2}\, m_{106}
 - p_1 + 2 p_{21} = 0 \,, \nonumber \\
 ({\rm \ref{Gam6}}) &\rightarrow&
 + 2\sqrt{2}\, m_{106} -p_{21} = 0 \,. \nonumber
\end{eqnarray}
These three equations fix $p_1$ and $p_{21}$, and set $y=0$
 (unless $a_1=0$, in which case $y$ remains arbitrary
 at this stage).
\par
The above calculation shows that any solution with $a_1\ne 0$
 will require
 the presence of $K_1$, and therefore the absence of an overall
 dilaton-dependent factor in front of the action.
\par
The result of the above calculation should be compared with the
 results
 presented in \cite{BedR89}.
 There the same terms that we consider above appeared in the
 Ansatz for the quartic action, and a similar calculation was done.
 The major difference is, however, that in
 \cite{BedR89} an $R^2$-action, related to the supersymmetrization
 of the Lorentz Chern-Simons terms, is present as well. Then the
 cancellation of the variation of the quartic action also involves
 contributions which arise iteratively from the quadratic and
 cubic action. One may check, that these contributions
 (which can be found in \cite{BedR89}) have the
 effect of setting $k_1=0$ and $y=-3$.
\par
The above calculation is a small part of the complete calculation
 which determines all coefficients in the Ansatz. But the
 general procedure should now be clear. The contributions to the
 variations are brought to a standard form, in such a way that
 the remaining structures are all independent.
 Of course one uses the identities mentioned in Table 2 to
 express the variation in terms of independent structures.
 For each independent structure in the variation of the action
 one finds an equation between the coefficients in the Ansatz.
 In solving the equations, free parameters may remain. Certainly
 one free parameter is associated with the normalisation of the
 action. Free parameters may also indicate that the Ansatz is
 overcomplete in the sense that a subset of the contributions
 to the Ansatz may be dependent.
 This occurs, for instance, for the seven terms in (\ref{L2}), of
 which
 only four are independent because of the identities
 (\ref{B-Identities}). Other free parameters indicate the presence
 of
 more than one solution to the problem of supersymmetrization.
 These aspects of our result will be discussed in the following
 section.
\par
Table 3 shows that the calculation splits in a natural way in
 two almost independent parts. The variations (A--D) and (I)
 (the $\psi$-sector)
 are independent of the dilatino $\lambda$, the variations
 (E--H) and (J) (the $\lambda$ sector)
 do depend on $\lambda$.
 All these $\lambda$-dependent variations come from
 $\lambda$-dependent terms in the Ansatz, except those
 due to the variation of the dilaton (see Table 1).
 The transformation 7 in Table 2 is only applied to a single
 sector of the Ansatz, (\ref{L4}), which does not contribute to the
 $\psi$-sector. Therefore it seems that, except for the variation
 of the dilaton factor in front of the total action, there is no
 contact between the $\psi$-sector and the $\lambda$-sector.
 However, the use of (\ref{2.20a}-\ref{2.22}) provides
 contributions which move from the $\psi$-sector to the
 $\lambda$-sector. Therefore it is essential
 to first work out the variations in the $\psi$-sector.
\par
 As we shall see, the equations resulting from the $\psi$-sector are
 very restrictive, and result in two independent solutions. The
 equations in the $\lambda$-sector are much less restrictive.
 As we discussed above, the cancellation of the
 (B)-variations involves only the identities
 (\ref{2.21}-\ref{2.22}).
 Using these, the (B)-variation produces
 $\bar\epsilon\,\lambda R^4$, (J)-terms.
 We expect the identities (\ref{2.20a}) and (\ref{2.20}) to
 play a role in the variation (A). Since (\ref{2.20a})
 contains a ${\cal D}\lambda$-contribution, the use of
 (\ref{2.20a}) in the cancellation of (A)
 provides a link between the $\psi$-sector
 and a variation containing ${\cal D}\lambda$.
 In Table 3 we see that there are several contributions to (A).
 Since no contractions between a $\Gamma$-matrix
 and the gravitino curvature are present in the
 Ansatz, only the variation of ${\cal L}_8$, the
 $\bar\psi\psi R^2{\cal D}R$-terms, can produce such a contraction.
 Therefore, all contributions containing ${\cal D}\lambda$ arising
 from the $\psi$-sector are proportional to the parameters in
 ${\cal L}_8$. However, the equations arising from the $\psi$-sector
 require, that all these parameters vanish!
\par
We conclude that the only link between the $\psi$- and
 $\lambda$-sector
 is through (B) and (J), and through the variation of $\phi^y$
 in front of the action, which also gives (J). Therefore we
 may choose a minimal option in the $\lambda$-sector, which is
 to include only those $\lambda$-dependent
 terms in the action which contribute to
 (J). As we see in Table 3, this is the sector ${\cal L}_{9}$,
 the $\bar\psi\psi_{(2)}R^3$-terms. Indeed, the calculation shows
 that cancellation of all $\lambda$-dependent terms in the
 variation can
 be achieved by including ${\cal L}_{9}$ only.
\par
Besides this minimal option we have also considered the inclusion
 of the sectors ${\cal L}_4,\ {\cal L}_{10-15}$. The
 variations from these terms have to cancel against each other. We
 have found that the resulting equations are not sufficiently
 restrictive to solve for all parameters in this part of the
 Ansatz. When discussing our results, in the next section,
 we will restrict ourselves to the minimal option mentioned
 above. Of course, this does not mean that we think that the
 coefficients in ${\cal L}_4,\ {\cal L}_{10-15}$
 are actually zero. It only
 means that these coefficients cannot be determined, in terms of
 a small number of free parameters, in the present calculation.
 The same remark holds for other sectors in the $R^4$-action
 which we have not considered in the construction of the
 Ansatz (see Section 3).

\newpage
\section{Results}
\vskip 12pt
Using the procedure discussed in the previous section,
 we find that supersymmetry requires that the bosonic terms
 must occur in the following combination:
\begin{eqnarray}
\label{5.1}
 {\cal L} & = & a_1 A_1 + (-16a_1+b)A_2+2a_1 A_3 +(12a_1-2b)A_4
 \nonumber \\
 & &+(-32a_1+4b)A_5 + (16a_1-2b)A_6+(-16a_1+2b)A_7
 \nonumber \\
 & &+ b_1 B_1 + b_2 B_2 + b_3 B_3 +
 (-{\textstyle{1\over 2}}b_2 - b_3 + 6\sqrt{2}\,b) B_4
 \nonumber \\
 & &+ 2b_2 B_5 + (b_1 + {\textstyle{1\over 4}}b_2
 + {\textstyle{1\over 2}}b_3 + 3\sqrt{2}\,b) B_6
 - (b_2 - 2b_3 - 12\sqrt{2}\, b)B_7
 \nonumber \\
 & & +{\textstyle{1\over 8}}\sqrt{2}\,a_1 K_1
 + {\textstyle{1\over 2}}\sqrt{2}\,(-a_1 +
 {\textstyle{1\over 8}}b) K_2 \,,
\end{eqnarray}
 where $b = {\textstyle{1\over 24}}\sqrt{2}\,(b_2 + 2b_3 + 2b_4)$.
 The coefficients $b_{1-4}$ remain free parameters after solving
 the equations. Three of these are redundant
 because
 of the three identities (\ref{B-Identities}),
 which imply that $B_{1-7}$ are not independent.
 We can therefore take arbitrary values for
 $b_{1-3}$, without changing the action.
 Thus $b$ and $a_1$ are the only true free
 parameters remaining in the action, which can therefore be written
 as a linear combination of two independent invariants.
\par
Expressed in terms of $X$, $Y_1$, $Y_2$ and $Z$
 the $R^4$-contribution in (\ref{5.1}) reads:
\begin{eqnarray}
 {\cal L}& = & cX+
 {\textstyle {{7!}\over{8}}}(a_1-{\textstyle{1\over 8}}b)Z
 +\big(6c-({\textstyle{1\over 8}}a_1+
 {\textstyle{3\over{64}}}b)\big)Y_1
 \nonumber \\
\label{5.2}
 &&\qquad
 +\big(-24c+{\textstyle{1\over 2}}(a_1
 -{\textstyle{1\over 8}}b)\big)Y_2 \, .
\end{eqnarray}
Here the coefficient $c$ is arbitrary and reflects the dependence
 of $X$, $Y_1$ and $Y_2$ discussed in Section 3.
\par
In (\ref{5.1}) we remark that for any nontrivial choice
 of $a_1$ and $b$ at least one of the $K$-terms is present.
 Our conclusion from Section 4, that the exponent $y$ in
 the factor $\phi^y$ must vanish, is therefore
 valid for arbitrary
 $a_1$ and $b$. Thus $a_1=0$ plays no special role in this respect.
\par
We will now discuss the two independent solutions contained in
 (\ref{5.1}). The first one is associated with $b=0$, the second
 with $b= 8a_1$. The most convenient way to express these
 two solutions in terms of $X$, $Z$, $Y_1$ and $Y_2$ is to
 take $c ={\textstyle{1\over {48}}}(a_1 -
 {\textstyle{1\over 8}}b)$ in (\ref{5.2}).
 The parameter $a_1$ is then a normalization factor, which we
 set equal to one.
\par
The complete action corresponding to the choice $b=0$
 (with $b_{1-3}=0$) is displayed
 in Appendix B (\ref{I1}).
 The bosonic part of this invariant reads:
\begin{eqnarray}
\label{5.3}
 I_1 & = & e\,\{ R_{abef}R_{abef}R_{cdgh}R_{cdgh}%
 -16R_{acef}R_{bcef}R_{adgh}R_{bdgh}%
 \nonumber \\
 & & + 2R_{abef}R_{cdef}R_{abgh}R_{cdgh}%
 + 12R_{abef}R_{cdef}R_{acgh}R_{bdgh}%
 \nonumber \\
 & & -32R_{abef}R_{cdef}R_{agch}R_{bgdh}%
 \nonumber \\
 & & +16R_{aebf}R_{cedf}R_{agbh}R_{cgdh}%
 -16R_{aebf}R_{cedf}R_{agch}R_{bgdh} \}%
 \nonumber \\
 & & +{\textstyle{1\over 8}}
 i\sqrt{2}\,\epsilon^{\mu_1\ldots\mu_{10}} B_{\mu_1\mu_2}
 R_{\mu_3\mu_4}{}^{ab}R_{\mu_5\mu_6}{}^{ab}
 R_{\mu_7\mu_8}{}^{cd}R_{\mu_9\mu_{10}}{}^{cd}
 \nonumber \\
 & & -{\textstyle{1\over 2}}
 i\sqrt{2}\,\epsilon^{\mu_1\ldots\mu_{10}} B_{\mu_1\mu_2}
 R_{\mu_3\mu_4}{}^{ab}R_{\mu_5\mu_6}{}^{ac}
 R_{\mu_7\mu_8}{}^{bd}R_{\mu_9\mu_{10}}{}^{cd}\,.
\end{eqnarray}
The $R^4$-terms in (\ref{5.3}) correspond to the combination
 ${\textstyle{1\over{48}}}\big(X + (6\times 7!)Z\big)$.
\par
Note that this
 solution has no terms linear in $H$. In \cite{GrSl87}
 it was found that in the string effective action the
 Riemann tensor should depend on the modified spin-connection
 $\Omega_-$ (see (\ref{2.6})). However, when $X$ and
 $Z$ are written in terms of the modified spin-connection
 $\Omega_-$, and one then expands in $H$, terms linear in
 $H$ cancel. Thus the effect of torsion
 appears only in the terms at least quadratic in $H$,
 which we do not consider here.
\par
The complete action corresponding to the choice $b=8a_1$
 (with $b_1=b_2=0$, $b_3=-48\sqrt{2}\,a_1$) is
 presented in (\ref{I2}).
 The bosonic part of this invariant is given by
\begin{eqnarray}
\label{5.4}
 I_2 & = & e\,\big\{ R_{abef}R_{abef}R_{cdgh}R_{cdgh}%
 -8 R_{acef}R_{bcef}R_{adgh}R_{bdgh}%
 \nonumber \\
 & & +2 R_{abef}R_{cdef}R_{abgh}R_{cdgh}%
 -4 R_{abef}R_{cdef}R_{acgh}R_{bdgh}%
 \nonumber \\
 & & + 96\sqrt{2}\,\hat H^{abc}\big(
 -{\textstyle{1\over 2}}R_{abem}R_{ghfm}{\cal D}_c R_{efgh}%
 + R_{abem}R_{ghfm}{\cal D}_e R_{cfgh} \big) \big\}%
 \nonumber \\
 & & +{\textstyle{1\over 8}}
 i\sqrt{2}\,\epsilon^{\mu_1\ldots\mu_{10}} B_{\mu_1\mu_2}
 R_{\mu_3\mu_4}{}^{ab}R_{\mu_5\mu_6}{}^{ab}
 R_{\mu_7\mu_8}{}^{cd}R_{\mu_9\mu_{10}}{}^{cd}\,.
\end{eqnarray}
The $R^4$-terms in (\ref{5.4}) are $-{\textstyle{1\over 2}} Y_1$.
The presence of $K_1$ implies that there is no
 factor $\phi^y$ in front of $I_2$.
\par
Using pair exchange for the
 Riemann tensor, all $R^4$-terms in (\ref{5.4})
 can be rewritten in terms of
\begin{eqnarray}
\label{Vdef}
 V_{\mu\nu\lambda\rho}&\equiv& R_{\mu\nu}{}^{ab}(\omega)
 R_{\lambda\rho}{}^{ab}(\omega)
\end{eqnarray}
and its contractions.
 Note
 that $V_{\mu\nu\lambda\rho}$ is the Lorentz-analogue of the
 Yang-Mills invariant
 ${\rm tr}\,F_{\mu\nu}F_{\lambda\rho}$.
 Because the connection $\Omega_-$
 transforms under supersymmetry as a Yang-Mills gauge field
 (compare (\ref{2.12}) and (\ref{2.25})), this
 analogy only holds if the spin connection in $V$ is
 $\Omega_-$. This suggests
 that the action should be rewritten in terms of the
 torsionful connection $\Omega_-$. Indeed, the two
 terms linear in $H$ in (\ref{5.4}) are precisely what is needed
 to introduce $H$-torsion, with the coefficient as in
 (\ref{2.6}), in the $R^4$-terms.
\par
The fermionic contributions to both $I_1$ and $I_2$ can be found
 in Appendix B. One surprise (for us) in this
 fermionic sector is that all terms of the type
 $\bar\psi\Gamma\psi R^2{\cal D} R$
 have a vanishing coefficient.
 Note that implicitly such terms appear in the
 action in (\ref{L1}) in the
 $\psi^2$-torsion in $\omega$, and in (\ref{L2}) in the
 supercovariantization in
 $\hat H$. Another way of presenting our result about
 ${\cal L}_8$ is to say that all such terms can be absorbed into
 $\psi^2$-torsion in $\omega$ and in supercovariantizations.
\par
Both the actions $I_1$ and $I_2$ contain terms dependent on the
 field $\lambda$. In Section 4 we discussed our procedure with
 respect to the $\lambda$-sector. Because of the vanishing of
 ${\cal L}_8$, it is possible to include only ${\cal L}_9$
 in the $\lambda$-sector, the so-called
 minimal option. All the coefficients $p_i$ are then
 determined.
\par
In the calculations leading to $I_1$ and $I_2$ we use the
 identities (\ref{2.21}) and (\ref{2.22}). The terms in the
 variation in which we encounter the left-hand-side of
 (\ref{2.21}) and (\ref{2.22}) are for $I_1$:
\begin{eqnarray}
\label{5.5}
 \bigl\{R_{abcd}R_{ajef}R_{bkgh}
 -{\textstyle{1\over 4}}R_{abcd}R_{abef}R_{jkgh}\bigr\}\,
 \bar\epsilon\,\Gamma_{cdefgh}\fcapslashcD \psi_{jk} &&
 \nonumber \\
 +\bigl\{2R_{abde}R_{abci}R_{fgcj}
 +12R_{acde}R_{afbi}R_{bgcj}\bigr\}\,
 \bar\epsilon\,\Gamma_{defg}\fcapslashcD \psi_{ij} &&
 \nonumber \\
 +\bigl\{-8R_{acbd}R_{aebh}R_{cfdi}
 +4R_{abcd}R_{abce}R_{dfhi}
 +4R_{bcad}R_{efah}R_{bcdi} &&
 \nonumber \\
 \phantom{123} +20R_{bcae}R_{adfh}R_{bcdi}
 +2R_{adef}R_{bcah}R_{bcdi}
 -16R_{abde}R_{cfah}R_{bcdi} &&
 \nonumber \\
 \phantom{123} +24R_{abde}R_{afch}R_{bcdi}
 -{\textstyle{1\over 2}}R_{abcd}R_{abcd}R_{efhi}
 +2R_{abce}R_{abdf}R_{cdhi} &&
 \nonumber \\
 \phantom{123}
 -R_{abcd}R_{abef}R_{cdhi}
 +8R_{acbe}R_{adbf}R_{cdhi}\bigr\}\,
 \bar\epsilon\,\Gamma_{ef}\fcapslashcD \psi_{hi} &&
 \nonumber \\
 -2R_{abde}R_{acfg}R_{bhci}\bar\epsilon\,\Gamma_{defgh}{\cal
 D}_{j}\psi_{ij} &&
 \nonumber \\
 +\bigl\{8R_{abce}R_{abdf}R_{cdgh}
 -4R_{acbe}R_{adbf}R_{cdgh}
 -8R_{abef}R_{acdg}R_{bcdh} &&
 \nonumber \\
 \phantom{12345}
 +8R_{adef}R_{bcag}R_{bcdh}\bigr\}\,
 \bar\epsilon\,\Gamma_{efg}{\cal D}_{i}\psi_{hi} &&
 \nonumber \\
 +\bigl\{32R_{acbe}R_{adbf}R_{cdeg}
 -20R_{abcd}R_{abef}R_{cdeg}&& \nonumber \\
 \phantom{12345}+20R_{abcd}R_{abce}R_{dfeg}\bigr\}\,
 \bar\epsilon\,\Gamma_{f}{\cal D}_{h}\psi_{gh}\,.&&
\end{eqnarray}
Using the identities (\ref{2.21}) and (\ref{2.22}) this can be
 expressed
 as derivatives of the equations of motion $\Psi_\mu$ and $\Lambda$
 of $\psi_\mu$ and $\lambda$, and
 terms proportional to $\psi R$ and $\lambda R$,
 which contribute to other variations. These last terms have been
 taken
 into account in the calculation.
 The equations of motion always occur in the
 combination $\Psi_\mu + {\textstyle{1\over 4}}\sqrt{2}\,
 \Gamma_\mu\Lambda$.
 The required additional variations of $\psi_\mu$
 and of $\lambda$, $\delta_\gamma\psi_\mu$ and
 $\delta_\gamma\lambda$,
 are given in (\ref{trans1}).
 Of course, the combination of the two equations of motion
 implies a relation between
 $\delta_\gamma\psi_\mu$ and $\delta_\gamma\lambda$. The fact that
 the
 only changes in the $\lambda$ transformation rules occur in
 this particular combination with $\delta_\gamma\psi_\mu$ is a
 consequence of the fact that we need only ${\cal L}_9$ in the
 $\lambda$-sector. The variation of ${\cal L}_9$ never gives
 rise to additional $\lambda$ equations of motion.
\par
For the invariant $I_2$ (\ref{5.4}) the remaining fermionic
 equations
 of motion arise from:
\begin{eqnarray}
\label{5.6}
 -{\textstyle{1\over 4}}R_{cdab}R_{efab}R_{ghjk}\,
 \bar\epsilon\,\Gamma_{cdefgh}\fcapslashcD \psi_{jk} &&
 \nonumber \\
 +\bigl\{2R_{ceab}R_{dfab}R_{cdhi}
 -{\textstyle{1\over 2}} R_{cdab}R_{cdab}R_{efhi}
 -R_{cdab}R_{efab}R_{cdhi} &&
 \nonumber \\
 \qquad
 + 4R_{cdab}R_{ceab}R_{dfhi}\bigr\}\,
 \bar\epsilon\,\Gamma_{ef}\fcapslashcD \psi_{hi} \,. &&
\end{eqnarray}
The corresponding modifications to the transformation rules
 of $\psi_\mu$ and $\lambda$ are given in (\ref{trans2}).
 The remaining variations containing
 bosonic equations of motion, which imply additional transformation
 rules for the bosonic fields, will not be presented explicitly.
 The new transformation rules of the bosonic fields are not
 immediately relevant for the compactification procedure.

Let us now come back in more detail to the analogy between these
 $R^4$-actions and quartic Yang-Mills invariants. We already
 mentioned above that the Riemann tensors in
 the bosonic part of $I_2$ can be expressed in
 terms of $V_{\mu\nu\lambda\rho}$ (\ref{Vdef}), if we use
 the torsionful spin connection $\Omega_-$. This requires
 the use of pair exchange for the Riemann tensor, which
 gives rise to additional $\bar\psi\Gamma\psi_{(2)}R^3$
 Noether terms, since:
\begin{eqnarray}
 R_{ab}{}^{cd}(\omega) & = & R^{\,cd}{}_{ab}(\omega) -
 {\textstyle{1\over 2}}\bar\psi^{[c}\Gamma^{d]}\psi_{ab}
 - \bar\psi^{[c}\Gamma_{[a}\psi^{d]}{}_{b]}
 \nonumber \\
 \label{pairex}
 &&\qquad
 +{\textstyle{1\over 2}}\bar\psi_{[a}\Gamma_{b]}\psi^{cd}
 + \bar\psi_{[a}\Gamma^{[c}\psi_{b]}{}^{d]} \,.
\end{eqnarray}
The additional fermionic terms due to pair exchange give
 contributions to the action which make it possible to write
 $I_2$ in terms of $V$ and
\begin{eqnarray}
\label{Wdef}
 W_{\mu\nu} &\equiv& R_{\mu\nu}{}^{ab}(\omega)\,\psi_{ab} \,.
\end{eqnarray}
$W$ is also
 the Lorentz form of a Yang-Mills invariant:
 ${\rm tr}\,F_{\mu\nu}\chi$.
\par
All contributions to (\ref{5.4})
 can be generalized to the $d=10$ Yang-Mills multiplet, if
 we replace in the action
\begin{eqnarray}
 V_{\mu\nu\lambda\rho} & \rightarrow &
 {\rm tr}\, F_{\mu\nu}(A)F_{\lambda\rho}(A) \,, \nonumber\\
\label{VWtoYM}
 W_{\mu\nu} &\rightarrow & {\rm tr}\,F_{\mu\nu}(A)\chi \,.
\end{eqnarray}
where $A_\mu$ and $\chi$ are the fields of the $d=10$ Yang-Mills
 multiplet. The resulting quartic Yang-Mills action will then
 be invariant under the transformations ({\ref{2.25}), (\ref{2.26}),
 if the Yang-Mills analogue of the terms (\ref{5.6}) allows the
 same treatment as in the case of the $R^4$-action.
 Writing (\ref{5.6})
 in terms of Yang-Mills fields we obtain:
\begin{eqnarray}
 \label{5.7}
 && -{\textstyle{1\over 4}}\,{\rm tr}\,F_{cd}F_{ef}
 \,
 \bar\epsilon\,\Gamma_{cdefgh}\,{\rm tr}\,F_{gh}\fcapslashcD \chi
 \nonumber \\
 && +(2\,{\rm tr}\,F_{ce}F_{df} -\,{\rm tr}\,F_{cd}F_{ef})
 \bar\epsilon\,\Gamma_{ef}\,{\rm tr}\,F_{cd}\fcapslashcD \chi
 \nonumber \\
 && \qquad
 -{\textstyle{1\over 2}}\,{\rm tr}\, F_{cd}F_{cd}
 \bar\epsilon\,\Gamma_{ef}\,{\rm tr}\,F_{ef}\fcapslashcD \chi
 + 4\,{\rm tr}\,F_{cd}F_{ce}
 \bar\epsilon\,\Gamma_{ef}\,{\rm tr}\,F_{df}\fcapslashcD \chi\,.
\end{eqnarray}
Now, the relevant
 terms in the
 $\chi$-equation of motion which follows from the {\it quadratic}
 Yang-Mills action read\footnote{We use here the form of the
 Yang-Mills-supergravity action given in \cite{BedR89}.}
\begin{eqnarray}
\label{eomchi}
 {\cal X} =
 e\phi^{-3}\{ \fcapslashcD (\omega,A)\chi
 +{\textstyle{1\over 4}}\Gamma^c\Gamma^{ab}\psi_c F_{ab}
 +{\textstyle{1\over 2}}\sqrt{2}\,\Gamma^{ab}F_{ab}\lambda\} \,,
\end{eqnarray}
so that the identity corresponding to (\ref{2.21}) is:
\begin{eqnarray}
\label{idchi}
 \fcapslashcD (\omega,A)\chi =
 e^{-1}\phi^3 {\cal X}
 -{\textstyle{1\over 4}}\Gamma^c\Gamma^{ab}\psi_c F_{ab}
 -{\textstyle{1\over 2}}\sqrt{2}\,\Gamma^{ab}F_{ab}\lambda \,.
\end{eqnarray}
So indeed we can express $\fcapslashcD \chi$ in terms of
${\cal X}$, and $\psi R$ and $\lambda R$ terms. Note that these
last terms take on exactly the same form as the $\psi R$ and
 $\lambda R$
contributions in (\ref{2.21}). This is of course essential for the
invariance of the quartic Yang-Mills action, since after the
use of the identity (\ref{idchi}) the rest of the calculation
should proceed in the same fashion as in the $R^4$-case.
\par
${\cal X}$ is the fermionic equation of motion of the
 $F^2$-action. Therefore, the ${\cal X}$ contributions in
 (\ref{5.7}) can only be cancelled by changing the $\chi$
 transformation rule if we include the supersymmetric $F^2$-action.
 In this way we obtain an action
\begin{eqnarray}
\label{YMaction}
 {\cal L} = R + \beta \,{\rm tr}\, F^2 + \gamma (\,{\rm tr}\,
 F^2)^2 \,,
\end{eqnarray}
and supersymmetry will require new transformation rules of
 $\chi$ and $A_\mu$ of order $\gamma/\beta$.
 As a byproduct of our analysis
 of $R^4$-actions we therefore find also the following
 Yang-Mills invariant (with $W_{\mu\nu} = \,{\rm
 tr}\,F_{\mu\nu}\chi$):
\begin{eqnarray}
\label{5.8}
{\cal L}_{\rm YM} & = & {\cal L}_R + {\cal L}_{F^2} \nonumber \\
&& +\gamma e\,\{-{\textstyle{1\over 2}}t^{\mu_1...\mu_8}
 \,{\rm tr}\,F_{\mu_1\mu_2}F_{\mu_3\mu_4}
 \,{\rm tr}\,F_{\mu_5\mu_6}F_{\mu_7\mu_8} \nonumber \\
&&+{\textstyle{i\over 8}}e^{-1}\sqrt{2}\,
 \epsilon^{\mu_1....\mu_{10}}B_{\mu_1\mu_2}
 \,{\rm tr}\,F_{\mu_3\mu_4}F_{\mu_5\mu_6}
 \,{\rm tr}\,F_{\mu_7\mu_8}F_{\mu_9\mu_{10}} \nonumber \\
&&+4\,\bar W^{\mu\nu}\Gamma^\lambda\,
 \,{\rm tr}\,\chi\,{\cal D}_\lambda F_{\mu\nu}
 -2\,{\rm tr}\,F^{\mu\nu}{\cal D}_\mu F^{\lambda\rho}\,
 \,{\rm tr}\,\bar\chi\,\Gamma_{\nu\lambda\rho}\chi \nonumber \\
&&-4\,\bar W^{\mu\nu}\Gamma^{\lambda\rho}{}_\nu\,{\rm tr}\,\chi\,
 {\cal D}_\mu F_{\lambda\rho} \nonumber \\
&&-8\,{\rm tr}\,F^{\mu\lambda}F^\nu{}_\lambda
 \,{\rm tr}\,\bar\chi\,\Gamma_\mu{\cal D}_\nu\chi
 -16\,\bar W^{\mu\nu}\Gamma_\nu
 \,{\rm tr}\,({\cal D}^\lambda\chi)F_{\mu\lambda}\}
 \nonumber \\
&& +\gamma e\sqrt{2}\,\biggl[
 \{ \,{\rm tr}\,F^{\mu\nu}F_{\mu\nu}
 \,\bar\lambda\,\Gamma_{\lambda\rho}
 -8 \,{\rm tr}\,F^{\mu\nu}F_{\mu\lambda}
 \,\bar\lambda\,\Gamma_{\nu\rho}
 -4 \,{\rm tr}\,F^{\mu}{}_{\lambda}F_{\nu\rho}
 \,\bar\lambda\,\Gamma_{\mu\nu} \nonumber \\
&&\qquad\quad
 +2 \,{\rm tr}\,F^{\mu\nu}F_{\lambda\rho}
 \,\bar\lambda\,\Gamma_{\mu\nu} \}\,W^{\lambda\rho}
 \nonumber \\
&&\quad +{\textstyle{1\over 2}}
 \,{\rm tr}\,F^{\mu\nu}F^{\sigma\tau}
 \,\bar\lambda\,\Gamma_{\mu\nu\sigma\tau\lambda\rho}\,
 W^{\lambda\rho} \biggr]
 \nonumber \\
&& + {\rm Noether\ terms} \,.
\end{eqnarray}
The complete invariant is presented in (\ref{LYM}), the $O(\gamma)$
 transformation rules of $\chi$ in (\ref{trans3}).
 In the above we have not considered the bosonic equations of motion
 nor the new transformation rules of $A_\mu$. We have checked that
 indeed the bosonic counterpart of (\ref{5.6}) also allows the
 generalization to an arbitrary Yang-Mills group.
\par
In the abelian case (\ref{5.8})
 reduces to the quartic contribution to the
Born-Infeld action \cite{BoIn34} coupled to supergravity, and
 agrees in the
 flat limit with
the globally supersymmetric Born-Infeld action presented in
 \cite{BeRaSe87}.
In the Yang-Mills case the structure of (\ref{5.8}) differs in the
flat limit from the result of \cite{BeRaSe87}, since in
 \cite{BeRaSe87} only
the symmetric Yang-Mills trace (i.e., $t^{\ldots}\,{\rm tr}\, F^4$)
 is
considered.
\par
The invariant $I_2$ corresponds to one particular choice of the
 coefficients $a_1$ and $b$ in (\ref{5.1}). One may wonder,
 whether other choices also lead to actions which have a
 Yang-Mills generalization. There are, for an arbitrary
 Yang-Mills group, eight independent $\,{\rm tr}\,F^4$ invariants.
 These are given by
\begin{eqnarray}
\label{YMstructures}
 {\rm YM}_1 & = & F_{\mu\nu}{}^I F^{\mu\nu\,J}
 F_{\lambda\rho}{}^{K}F^{\lambda\rho\,L} \,,
 \nonumber \\
 {\rm YM}_2 & = & F_{\mu\nu}{}^I F^{\nu}{}_{\lambda}{}^J
 F^{\mu\rho\,K}F_{\rho\lambda}{}^L \,,
 \nonumber \\
 {\rm YM}_3 & = & F_{\mu\nu}{}^I F^{\nu}{}_{\lambda}{}^J
 F_{\lambda\rho}^{K}F_{\rho}{}^{\mu\,L} \,,
 \nonumber \\
 {\rm YM}_4 & = & F_{\mu\nu}{}^I F_{\lambda\rho}{}^J
 F^{\mu\nu\,K}F^{\lambda\rho\,L} \,,
 \nonumber
\end{eqnarray}
multiplied by either $\,{\rm tr}\,T_IT_J\,{\rm tr}\,T_KT_L$, giving
 ${\rm YM}_i(1)$, or
 $\,{\rm tr}\,T_IT_JT_KT_L$, giving ${\rm YM}_i(2)$.
 Here $T_I$ are the Yang-Mills generators in the
 fundamental representation.
 These eight possibilities give the following $R^4$-actions
 if we work them out for the $SO(9,1)$ Lorentz group\footnote{In
 this
 calculation we use pair exchange and the
 cyclic identity for the Riemann tensor.}:
\begin{eqnarray}
\label{YMcorrespondence}
 {\rm YM}_1(1) \quad\rightarrow\quad A_1 \,, &&
 {\rm YM}_1(2) \quad\rightarrow\quad A_2 \,, \nonumber \\
 {\rm YM}_2(1) \quad\rightarrow\quad A_2 \,, &&
 {\rm YM}_2(2) \quad\rightarrow\quad
 {\textstyle{1\over 4}}A_4 - A_5 + A_7 \,, \nonumber \\
 {\rm YM}_3(1) \quad\rightarrow\quad A_2 \,, &&
 {\rm YM}_3(2) \quad\rightarrow\quad A_6 \,, \nonumber \\
 {\rm YM}_4(1) \quad\rightarrow\quad A_3 \,, &&
 {\rm YM}_4(2) \quad\rightarrow\quad A_4 \,.
\end{eqnarray}
Note that $Z$ (\ref{3.6}) has the wrong combination of
 $A_5$ and $A_7$ to be the Lorentz-case of a general
 Yang-Mills invariant. The only way to avoid having
 $Z$ in our solution (\ref{5.1})
 is to
 choose $b=8a_1$, which leads to $I_2$.
 Thus $I_{{\rm YM}}$ is the only Yang-Mills invariant which we can
 reconstruct from our result. This implies
 that a supersymmetric action of the type $t^{\ldots}\,{\rm
 tr}\,F^4$, which
 would correspond to the generalization of $Y_2$, does
 not exist for arbitrary Yang-Mills groups.
\par
The action (\ref{5.8}) can be generalized in the following way. We
 may choose a semi-simple gauge group of of the form
 $G\times SO(9,1)$.
 Then we can identify the gauge field of $SO(9,1)$ with
 $\Omega_-$, and the corresponding field strength with the Riemann
 tensor. The invariant (\ref{YMaction}) then takes on the form
\begin{eqnarray}
\label{RFaction}
 {\cal L} & = & R + \beta\,{\rm tr}\,F^2 +
 \gamma (R^2 + \,{\rm tr}\,F^2)^2 \,.
\end{eqnarray}
Note that an $R^2$-term is not required for invariance. In the
 absence of quadratic terms invariance holds up to (\ref{5.7})
 for $G\times SO(9,1)$. For the contributions containing
 $\fcapslashcD \chi$, where $\chi$ is the partner of the
 $G$-gauge field, we use (\ref{idchi}). This
 requires the presence of the standard $F^2$-action. For the
 contributions containing $\fcapslashcD \psi_{ab}$
 we use (\ref{2.21}), which contains an equation of motion of
 the $R$-action. Therefore no $R^2$-action is needed to
 cancel these particular variations.

\newpage
\section{Discussion}

In this paper, we have found that two supersymmetric invariants of
 the type $R+\gamma R^4$ exist. As a by-product, we have also
 obtained
 the leading terms of a locally supersymmetric
 $\,{\rm tr}\, F^2+\gamma (\,{\rm tr}\, F^2)^2$-invariant.
\par
Let us now compare our results to the effective action
 obtained by other methods. The tree-level string amplitude
 contributions to ${\cal L}_{\rm eff}$
 contain the action ${\cal L}_R$, (\ref{2.14}),
 with the Yang-Mills contribution ${\cal L}_{F^2}$. The
 field strength $H$ of the antisymmetric tensor gauge field
 $B_{\mu\nu}$
 is modified with Yang-Mills and Lorentz Chern-Simons terms.
 As discussed in Section 2, supersymmetry requires the
 presence of ${\cal L}_{R^2}$ terms, and quartic contributions of
 the form
 $(R^2 + \,{\rm tr}\,F^2)^2$. In these quadratic and quartic
 actions the
 Riemann tensor depends on $\Omega_-$, and the couplings to the
 dilaton are limited to the same overall factor $\phi^{-3}$ which is
 also present in ${\cal L}_R$. As we discussed in Section 4, this
 action does not contain a term $K_1$ (\ref{K1}), so that the
 overall factor $\phi^{-3}$ does not interfere with the
 $B_{\mu\nu}$-gauge transformations.
 The result of supersymmetrizing the Lorentz
 Chern-Simons terms \cite{BedR89} agrees (up to field redefinitions)
 with the determination of the bosonic part by a string amplitude
 calculation \cite{GrSl87}.
\par
In \cite{GrSl87} a different basis is used for the independent
 fields. The dilaton is denoted by the field $D$, with the
 correspondence $\phi = \exp({\textstyle{2\over 3}}\sqrt{2}\,D)$,
 $\phi$ being our scalar field. The tenbein in \cite{GrSl87}
 differs by a factor $\phi^{-3/8}$ from our tenbein. With this
 rescaling, we find indeed that the modified Riemann tensor $\bar
 R$ in
 \cite{GrSl87}, which contains ${\rm e}^{D}H$ and ${\cal D}{\cal
 D}D$
 contributions, becomes equal to $R_{\mu\nu}{}^{ab}(\Omega_-)$.
\par
Among the tree level terms obtained in \cite{GrSl87} is also the
 contribution $\zeta(3)X$, with $X$ given in (\ref{3.6}). After the
 rescaling mentioned above, this term also obtains the overall
 factor
 $\phi^{-3}$. Therefore we must conclude from our analysis, that
 this term does not have a supersymmetric completion. As we have
 seen, the supersymmetrization of $X$ requires the presence of
 both $K_1$ and $K_2$ (\ref{K1}, \ref{K2}),
 which because of $B$-gauge invariance
 conflicts with the presence of the
 $\phi^{-3}$-factor\footnote{The terms $\phi^{-3}K_i$ are
 gauge invariant under modified $B$-gauge transformations:
 $\delta B_{\mu\nu} = 2 (\partial_{[\mu}\Lambda_{\nu]}
 + 3(\phi^{-1}\partial_{[\mu}\phi)\,\Lambda_{\nu]})$. However,
 the conflict is now shifted to the $H$-dependence in ${\cal L}_R$.
 The field strength $H$ has to be modified to be invariant
 under the new $B$-gauge transformations. This
 breaks the supersymmetry of ${\cal L}_R$. These modified gauge
 transformations are discussed in \cite{CLNY88}.}.
 Therefore we still do not understand the properties of $\zeta(3)X$
 in relation to supersymmetry in ten dimensions.
\par
At the one-loop level string amplitudes reveal again the presence
 of the $X$-term, as well as further $(R^2+\,{\rm tr}\,F^2)^2$-terms
 \cite{ElJeMiz87}, \cite{AbKuSa88a}.
 However, the one-loop
 contributions to ${\cal L}_{\rm eff}$ have {\em no}
 overall dilaton factor. One also finds a contribution proportional
 to $\,{\rm tr}\,F^4$. For $E_8\times E_8$ this term can be
 rewritten in the form $(\,{\rm tr}\,F^2)^2$, but this is
 not possible for $SO(32)$.
\par
Comparing now to our results in Section 5, we see that we can
 indeed
 supersymmetrize the one-loop contributions to the effective
 action,
 except for $\,{\rm tr}\,F^4$, which remains a problem in case the
 gauge group is $SO(32)$. In Section 5 we showed
 that the supersymmetrization of $(R^2 + \,{\rm tr}\,F^2)^2$
 requires an $F^2$-contribution to the action, but no $R^2$-terms.
 This implies that the $R^2$-contributions to the effective action
 are completely determined by the supersymmetrization of the
 Lorentz Chern-Simons terms, or, in string amplitude terminology, by
 the tree-level contributions.
\par
The counterterms required for the cancellation
 of anomalies for the gauge group $E_8\times E_8$ are,
 schematically,
 \cite{GS84}
\begin{eqnarray}
 {\cal L}_{\rm counter} & \sim &
 \epsilon^{\mu_1\ldots\mu_{10}} B_{\mu_1\mu_2} \{
 \,{\rm tr}\,R^4 + {\textstyle{1\over 4}}(\,{\rm tr}\,R^2)^2
 + (\,{\rm tr}\,R^2)(\,{\rm tr}\,F^2) +
 (\,{\rm tr}\,F^2)^2 \}_{\mu_3\ldots\mu_{10}} \,.
 \nonumber
\end{eqnarray}
All these counterterms can be seen as part of the supersymmetric
 actions presented in Section 5. Note in particular, that
 we also obtain the relative coefficient ${\textstyle{1\over 4}}$
 between the two $R^4$-terms. Thus we find that these counterterms
 are indeed linked by supersymmetry to the known bosonic one-loop
 contributions to the quartic effective string action.
 The other counterterms presented
 in \cite{GS84}, which contain products of
 Chern-Simons forms, belong in our terminology to actions
 $R^n$ with $n>4$.

In a recent paper by Duff and Lu \cite{DuLu91} it was argued that
 the
 coupling of the heterotic five-brane \cite{St90} $\sigma$-model to
 background supergravity fields implies the existence of
 quartic terms in the Riemann tensor and Yang-Mills field
 strength. However, these are obtained in the
 version of $N=1,\ d=10$ supergravity with
 a six-index antisymmetric gauge field, which is
 related to our $B_{\mu\nu}$ by a duality transformation. Let us
 therefore consider the effect of a duality
 transformation on the quartic action we obtain in this paper.
\par
For this duality transformation we focus again on the
 $B\wedge R\wedge R\wedge R\wedge R$ terms
 . They are related to
 Chern-Simons forms. The usual Lorentz Chern-Simons term
 $\omega_3$ appears as
 a modification to the field strength $H$ of the gauge field $B$,
 schematically, this reads: $H\sim \partial
 B + \,{\rm tr}\, (\omega\wedge \partial \omega +
 \omega\wedge \omega\wedge \omega)$, along
 with the Yang-Mills Chern-Simons term \cite{ChapMan83}.
 In the dual version of $d=10$ supergravity with a
 six-index gauge field
 Chern-Simons terms are absent,
 but are replaced by an interaction term of the form $A_{(6)}\wedge
 R\wedge R$ in the action.
\par
By a similar duality transformation, the terms $B\wedge R\wedge
 R\wedge R\wedge R$ will give rise to the Chern-Simons forms
 $\omega_7$,
\begin{eqnarray}
 H_{(7)} &\sim &
 \partial A_{(6)} + {\rm tr}\,(
 \omega\wedge\partial\omega\wedge\partial\omega\wedge
 \partial\omega) +\ldots \,,
 \nonumber
\end{eqnarray}
in the seven-index field strength of $A_{(6)}$ in the six-index
 version of $d=10$ supergravity. Such terms are indeed required in
 the anomaly cancellations in the six-index version \cite{GaNi85}.
\par
In this paper we have supersymmetrized the one-loop, quartic
 terms which appear in the bosonic
 string effective action. We do not find a
 supersymmetric completion for
 the $\zeta(3)\phi^{-3}X$-term, which is part of the tree-level
 effective action. This failure may be due to the fact that we
 limited ourselves to the use of the physical fields of $d=10,\ N=1$
 supergravity. Failure of the Noether method may of course indicate
 the necessity of introducing additional fields. These could
 correspond to massive fields, perhaps related to auxiliary fields
 of the $d=10,\ N=1$ supergravity multiplet, which become
 propagating fields in the higher derivative actions we
 have considered.

\section*{Acknowledgements}

We thank Eric Bergshoeff for a number of useful discussions. This
 work
 is financially supported by the Stichting voor Fundamenteel
 Onderzoek
 der Materie (F.O.M.). One of us (A.W.) gratefully acknowledges the
 financial support of the Deutsche Forschungs Gemeinschaft (DFG)
 under contract nr. Wi 1033/1-1.

\newpage
\appendix
\section{Appendix A}
This Appendix is devoted to the presentation of the various sectors
 we
 constructed for the Ansatz.
 We write the Ansatz
 as the sum (\ref{Ltot}):
\begin{eqnarray}
{\cal L}_{\rm tot} & = & \gamma\,e\phi^{y}\sum_i{\cal L}_i \,.
 \nonumber
\end{eqnarray}
\par
The first purely bosonic sector is formed by seven terms of the
 form $R^4$ and
 therefore contains all possible independent contractions of four
 Riemann
 tensors:
\begin{eqnarray}
\label{L1}
{\cal L}_1 & = &
 +a_{1}R_{abcd}R_{abcd}R_{efgh}R_{efgh}
 +a_{2}R_{abcd}R_{abce}R_{dfgh}R_{efgh} \nonumber \\
&& +a_{3}R_{abcd}R_{abef}R_{cdgh}R_{efgh}
 +a_{4}R_{abce}R_{abdf}R_{cdgh}R_{efgh} \nonumber \\
&& +a_{5}R_{abce}R_{abdg}R_{cfdh}R_{efgh}
 +a_{6}R_{acbd}R_{aebg}R_{cfdh}R_{efgh} \nonumber \\
&& +a_{7}R_{acbe}R_{adbg}R_{cfdh}R_{efgh}\,.
\end{eqnarray}
\par
The second sector in our Ansatz consists of seven terms of type
$HR^2{\cal D}R$. Its explicit form is:
\begin{eqnarray}
\label{L2}
 {\cal L}_2 & = &
 +b_{1}H_{aef}R_{adbc}R_{ghbc}{\cal D}_dR_{efgh}
 -b_{2}H_{aef}R_{abcg}R_{bdch}{\cal D}_dR_{efgh} \nonumber \\
&& +b_{3}H_{abd}R_{abce}R_{ghcf}{\cal D}_dR_{efgh}
 +b_{4}H_{abe}R_{abcd}R_{ghcf}{\cal D}_dR_{efgh} \nonumber \\
&& +b_{5}H_{abe}R_{adcg}R_{bfch}{\cal D}_dR_{efgh}
 +b_{6}H_{abc}R_{abef}R_{cdgh}{\cal D}_dR_{efgh} \nonumber \\
&& +b_{7}H_{abe}R_{adcf}R_{bcgh}{\cal D}_dR_{efgh} \,.
\end{eqnarray}
It is important to realize that the seven terms in this sector are
overcomplete. This is due to the fact that the Bianchi identity for
 the
$H$-field implies the following relations among the different terms:
\begin{eqnarray}
\label{B-Identities}
 0 & = & {D}_{[a}\hat H_{bcd]}
 R_{abef}R_{cdgh}R_{efgh} = B_1+B_6 \,,
 \nonumber \\
 0 & = & {D}_{[a}\hat H_{bcd]}
 R_{abef}R_{cegh}R_{dfgh} =
 {\textstyle{1\over 4}}B_1
 -{\textstyle{1\over 2}}B_3
 +{\textstyle{1\over 2}}B_4+B_7\,, \nonumber\\
 0 & = & {D}_{[a}\hat H_{bcd]}R_{abef}R_{cgeh}R_{dgfh} =
 {\textstyle{1\over 2}}B_2
 -{\textstyle{1\over 4}}B_3+B_5 \,.
\end{eqnarray}
The latter results are obtained by performing a partial
 integration. Note that
 these identities are valid modulo terms of the form
 $\bar\psi\Gamma\psi_{(2)}R^3$. This is related to the fact that the
 Bianchi-identity of the $H$-field involves a supercovariant
 derivative.
 The equations (\ref{B-Identities}) imply that three of the
 coefficients $b_i$
 can be chosen arbitrarily.
\par
The third purely bosonic sector consists of two terms of type
 $BR^4$:
\begin{eqnarray}
\label{L3}
{\cal L}_3 & = &
 +k_{1}ie^{-1}\epsilon_{abcdefijkl}
 B_{ab}R_{cdgh}R_{efgh}R_{ijmn}R_{klmn}
 \nonumber \\
&&
 +k_{2}ie^{-1}\epsilon_{abcdefhikl}
 B_{ab}R_{cdgj}R_{efgm}R_{hijn}R_{klmn}\,.
\end{eqnarray}
There are four terms of the structure
 $(\phi^{-1}\partial\phi)R^2{\cal D}R$:
\begin{eqnarray}
\label{L4}
{\cal L}_4 & = &
 + \phi^{-1}\partial_a\phi\, (
 c_1 R_{bcde}R_{fgde}{\cal D}_aR_{bcfg}
 +c_2 R_{bcde}R_{cfeg}{\cal D}_aR_{bfdg}
 \nonumber \\
 & & -c_3 R_{abcd}R_{defg}{\cal D}_bR_{cefg}
 -c_4 R_{abcd}R_{defg}{\cal D}_cR_{befg})
 \,.
\end{eqnarray}
This completes the list of the purely bosonic sectors.
\par
We considered 17 terms of the type
$\bar\psi_{(2)}\Gamma\psi_{(2)}R{\cal D}R$:
\begin{eqnarray}
\label{L5}
 {\cal L}_5
 & = &
 +\{d_{1}R_{bcae}{\cal D}_aR_{bcdf}
 +d_{2}R_{bcad}{\cal D}_aR_{bcef}
 +d_{3}R_{abcd}{\cal D}_eR_{abcf}\}\,
 \bar\psi_{dg}\Gamma_{e}\psi_{fg} \nonumber \\
&& +\{d_{5}R_{cdab}{\cal D}_eR_{fgab}
 +d_{6}R_{ceab}{\cal D}_dR_{fgab}
 +d_{7}R_{cfab}{\cal D}_dR_{egab} \nonumber \\
&&\qquad
 +d_{8}R_{acbf}{\cal D}_eR_{adbg}
 +d_{9}R_{acbf}{\cal D}_dR_{aebg} \nonumber \\
&&\qquad
 +d_{10}R_{aebf}{\cal D}_dR_{acbg}\}\,
 \bar\psi_{cd}\Gamma_{e}\psi_{fg} \nonumber \\
&& +d_{21}R_{bcaf}{\cal D}_aR_{bcde}
 \,\bar\psi_{gh}\Gamma_{def}\psi_{gh} \nonumber \\
&& +\{d_{22}R_{deab}{\cal D}_cR_{fgab}
 +d_{23}R_{dgab}{\cal D}_cR_{efab}
 +d_{24}R_{acbd}{\cal D}_fR_{aebg}\}\,
 \bar\psi_{ch}\Gamma_{def}\psi_{gh} \nonumber \\
&& +\{-d_{25}R_{efac}{\cal D}_aR_{bdgh}
 -d_{26}R_{ghaf}{\cal D}_aR_{bcde}
 -d_{27}R_{chaf}{\cal D}_aR_{bgde} \nonumber \\
&&\qquad
 -d_{28}R_{efac}{\cal D}_bR_{adgh}
 +d_{29}R_{cfah}{\cal D}_bR_{agde}\}\,
 \bar\psi_{bc}\Gamma_{def}\psi_{gh} \,.
\end{eqnarray}
Next, there are six terms with the structure
$\bar\psi_{(2)}\Gamma{\cal D}\psi_{(2)}R^2$:
\begin{eqnarray}
\label{L6}
{\cal L}_6
 & = &
 +f_{1}R_{adbc}R_{aebc}
 \,\bar\psi_{fg}\Gamma_{d}{\cal D}_e\psi_{fg}
 +f_{2}R_{acbd}R_{aebf}
 \,\bar\psi_{cg}\Gamma_{d}{\cal D}_e\psi_{fg} \nonumber \\
&& +f_{3}R_{bcad}R_{fgae}
 \,\bar\psi_{bc}\Gamma_{d}{\cal D}_e\psi_{fg}
 +f_{4}R_{bcaf}R_{deag}
 \,\bar\psi_{bc}\Gamma_{d}{\cal D}_e\psi_{fg} \nonumber \\
&& -f_{5}R_{cdaf}R_{egab}
 \,\bar\psi_{bh}\Gamma_{cde}{\cal D}_f\psi_{gh}
 +f_{6}R_{cdab}R_{efgh}
 \,\bar\psi_{ab}\Gamma_{cde}{\cal D}_f\psi_{gh}\,.
\end{eqnarray}
For the Noether sector, the terms of type
 $\bar\psi\Gamma\psi_{(2)}R^3$,
 we constructed 92 independent terms:
\begin{eqnarray}
\label{L7}
 {\cal L}_7
 & = &
 +\{m_{1}R_{afbg}R_{acde}R_{bcde}
 +m_{2}R_{afbc}R_{agde}R_{bcde} \nonumber \\
&&\qquad
 +m_{3}R_{bfad}R_{cgae}R_{bcde}\}\,
 \bar\psi_{h}\Gamma_{f}\psi_{gh} \nonumber \\
&& +\{m_{4}R_{efgh}R_{abcd}R_{abcd}
 +m_{5}R_{efag}R_{bhcd}R_{abcd}
 +m_{6}R_{egaf}R_{bhcd}R_{abcd} \nonumber \\
&&\qquad
 +m_{7}R_{ghae}R_{bfcd}R_{abcd}
 +m_{8}R_{ghaf}R_{becd}R_{abcd}
 +m_{9}R_{efab}R_{ghcd}R_{abcd} \nonumber \\
&&\qquad
 +m_{10}R_{egab}R_{fhcd}R_{abcd}
 +m_{11}R_{efab}R_{agcd}R_{bhcd}
 +m_{12}R_{egab}R_{afcd}R_{bhcd} \nonumber \\
&&\qquad
 +m_{13}R_{fgab}R_{aecd}R_{bhcd}
 +m_{14}R_{ghab}R_{aecd}R_{bfcd}
 +m_{15}R_{efac}R_{bgad}R_{bhcd} \nonumber \\
&&\qquad
 +m_{16}R_{egac}R_{bfad}R_{bhcd}
 +m_{17}R_{fgac}R_{bead}R_{bhcd}
 +m_{18}R_{ghac}R_{bead}R_{bfcd} \nonumber \\
&&\qquad
 +m_{19}R_{aecg}R_{bfdh}R_{abcd}
 +m_{20}R_{aecg}R_{bfad}R_{bhcd}
 +m_{21}R_{afcg}R_{bead}R_{bhcd} \nonumber \\
&&\qquad
 +m_{22}R_{aebg}R_{afcd}R_{bhcd}
 +m_{23}R_{afbg}R_{aecd}R_{bhcd}\}\,
 \bar\psi_{e}\Gamma_{f}\psi_{gh} \nonumber \\
&& +\{m_{30}R_{efhi}R_{abcd}R_{abcd}
 +m_{31}R_{efah}R_{bicd}R_{abcd}
 +m_{32}R_{hiae}R_{bfcd}R_{abcd} \nonumber \\
&&\qquad
 +m_{34}R_{efab}R_{hicd}R_{abcd}
 +m_{35}R_{ehab}R_{ficd}R_{abcd}
 +m_{37}R_{efab}R_{ahcd}R_{bicd} \nonumber \\
&&\qquad
 +m_{38}R_{ehab}R_{afcd}R_{bicd}
 +m_{39}R_{hiab}R_{aecd}R_{bfcd}
 +m_{42}R_{efac}R_{bhad}R_{bicd} \nonumber \\
&&\qquad
 +m_{43}R_{ehac}R_{bfad}R_{bicd}
 +m_{44}R_{hiac}R_{bead}R_{bfcd}
 +m_{47}R_{aech}R_{bfdi}R_{abcd} \nonumber \\
&&\qquad
 +m_{48}R_{aech}R_{bfad}R_{bicd}
 +m_{49}R_{aebh}R_{ficd}R_{abcd}\}\,
 \bar\psi_{g}\Gamma_{efg}\psi_{hi} \nonumber \\
&& +\{m_{33}R_{efah}R_{bgcd}R_{abcd}
 +m_{36}R_{efab}R_{ghcd}R_{abcd}
 +m_{40}R_{efab}R_{agcd}R_{bhcd} \nonumber \\
&&\qquad
 +m_{41}R_{ehab}R_{afcd}R_{bgcd}
 +m_{45}R_{efac}R_{bgad}R_{bhcd} \nonumber \\
&&\qquad
 +m_{46}R_{ehac}R_{bfad}R_{bgcd}\}\,
 \bar\psi_{i}\Gamma_{efg}\psi_{hi} \nonumber \\
&& +\{m_{501}R_{dhef}R_{agbc}R_{aibc}
 +m_{502}R_{efhi}R_{adbc}R_{agbc}
 +m_{51}R_{efad}R_{hibc}R_{agbc} \nonumber \\
&&\qquad
 +m_{52}R_{efad}R_{ghbc}R_{aibc}
 +m_{53}R_{efah}R_{dgbc}R_{aibc}
 +m_{54}R_{efah}R_{gibc}R_{adbc} \nonumber \\
&&\qquad
 +m_{55}R_{efah}R_{dibc}R_{agbc}
 +m_{56}R_{deah}R_{fgbc}R_{aibc}
 -m_{57}R_{deah}R_{fibc}R_{agbc} \nonumber \\
&&\qquad
 +m_{58}R_{ehad}R_{fgbc}R_{aibc}
 -m_{59}R_{ehad}R_{fibc}R_{agbc}
 +m_{60}R_{hiae}R_{dfbc}R_{agbc} \nonumber \\
&&\qquad
 +m_{61}R_{hiae}R_{fgbc}R_{adbc}
 +m_{62}R_{hiad}R_{efbc}R_{agbc}
 +m_{63}R_{efbd}R_{agch}R_{aibc} \nonumber \\
&&\qquad
 +m_{64}R_{efbh}R_{adcg}R_{aibc}
 +m_{65}R_{efbh}R_{agci}R_{adbc}
 +m_{66}R_{efbh}R_{adci}R_{agbc} \nonumber \\
&&\qquad
 +m_{67}R_{debh}R_{aicf}R_{agbc}
 +m_{68}R_{ehbd}R_{aicf}R_{agbc}
 +m_{69}R_{hibe}R_{adcf}R_{agbc} \nonumber \\
&&\qquad
 -m_{70}R_{deab}R_{fgac}R_{hibc}
 -m_{71}R_{dhab}R_{efac}R_{gibc}
 -m_{72}R_{efab}R_{hiac}R_{bdcg} \nonumber \\
&&\qquad
 -m_{73}R_{ehab}R_{fiac}R_{bdcg}
 -m_{74}R_{efab}R_{ghac}R_{bdci}
 -m_{75}R_{efab}R_{dhac}R_{bgci} \nonumber \\
&&\qquad
 -m_{76}R_{deab}R_{fhac}R_{bgci}
 +m_{77}R_{efab}R_{cdah}R_{bgci}
 +m_{78}R_{ehab}R_{cdaf}R_{bgci} \nonumber \\
&&\qquad
 +m_{79}R_{adbe}R_{cfah}R_{bgci}\}\,
 \bar\psi_{d}\Gamma_{efg}\psi_{hi} \nonumber \\
&& +\{m_{90}R_{deai}R_{fgbc}R_{ahbc}
 -m_{91}R_{deab}R_{fgac}R_{bhci}\}\,
 \bar\psi_{j}\Gamma_{defgh}\psi_{ij} \nonumber \\
&& +\{m_{92}R_{deai}R_{fgbc}R_{ajbc}
 +m_{93}R_{ijad}R_{efbc}R_{agbc}
 +m_{94}R_{deai}R_{fjbc}R_{agbc} \nonumber \\
&&\qquad
 +m_{95}R_{debi}R_{afcj}R_{agbc}
 -m_{96}R_{deab}R_{fiac}R_{bgcj}\}\,
 \bar\psi_{h}\Gamma_{defgh}\psi_{ij} \nonumber \\
&& +\{m_{97}R_{deci}R_{fgab}R_{hjab}
 +m_{98}R_{cdij}R_{efab}R_{ghab}
 +m_{981}R_{deij}R_{cfab}R_{ghab} \nonumber \\
&&\qquad
 +m_{99}R_{deai}R_{fgbj}R_{chab}
 +m_{100}R_{deac}R_{ijbf}R_{ghab}
 +m_{101}R_{diac}R_{efbj}R_{ghab} \nonumber \\
&&\qquad
 +m_{102}R_{cdai}R_{efbj}R_{ghab}
 +m_{103}R_{deac}R_{fgbi}R_{hjab} \nonumber \\
&&\qquad
 +m_{104}R_{deac}R_{fgbi}R_{ahbj}\}\,
 \bar\psi_{c}\Gamma_{defgh}\psi_{ij} \nonumber \\
&& +\{m_{105}R_{cdaj}R_{efbk}R_{ghab}
 +m_{106}R_{cdjk}R_{efab}R_{ghab}\}\,
 \bar\psi_{i}\Gamma_{cdefghi}\psi_{jk}\,.
\end{eqnarray}
Finally, in the cancellation mechanism we also included:
\begin{eqnarray}
\label{L8}
 {\cal L}_{8} & : & {\rm Terms \ of \ type }\phantom{AB}\bar\psi
 \Gamma^{(1)}\psi\,R^2{\cal D}R \nonumber \\
 & & {\rm and }\phantom{AB} \bar\psi
 \Gamma^{(5)}\psi\,R^2{\cal D}R \,.
\end{eqnarray}
Altogether there are 70 terms of this type.
 In our solutions we
 find that all these terms have to vanish. We will therefore
 not write them explicitly.

\par
In principle there are 19 additional sectors to be included in the
 Ansatz. Roughly speaking, these have fewer fields and
 more derivatives.
 These sectors consist of the following structures:
\begin{eqnarray}
 (\bar\psi\psi)R{\cal D}^3R\,; \quad
 (\bar\psi\psi){\cal D}R{\cal D}^2R \,; \quad
 (\bar\psi\psi_{(2)})R{\cal D}^2R \,; \quad
 (\bar\psi\psi_{(2)})({\cal D}R)^2 \,;
 \nonumber \\
 (\bar\psi\psi){\cal D}^5R \,;\quad
 (\bar\psi\psi_{(2)}){\cal D}^4R \,;\quad
 (\bar\psi_{(2)}\psi_{(2)}){\cal D}^3R \,;\quad
 (\bar\psi_{(2)}{\cal D}\psi_{(2)}){\cal D}^2R \,; \nonumber \\
 ({\cal D}\bar\psi_{(2)}{\cal D}\psi_{(2)}){\cal D}R \,;\quad
 (\bar\psi_{(2)}{\cal D}^2\psi_{(2)}){\cal D}R \,;\quad
 ({\cal D}\bar\psi_{(2)}{\cal D}^2\psi_{(2)})R \,;\quad
 (\bar\psi_{(2)}{\cal D}^3\psi_{(2)})R \,; \nonumber \\
 (\bar\psi{\cal D}\psi_{(2)})R{\cal D}R \,;\quad
 (\bar\psi{\cal D}^2\psi_{(2)})R^2 \,;\quad
 (\bar\psi{\cal D}\psi_{(2)}){\cal D}^3R \,;\quad
 (\bar\psi{\cal D}^2\psi_{(2)}){\cal D}^2R \,; \nonumber \\
 (\bar\psi{\cal D}^3\psi_{(2)}){\cal D}R \,;\quad
 (\bar\psi\psi_{(2)}){\cal D}^4R \,;\quad
 \hat H{\cal D}R{\cal D}^2R\,.
 \nonumber
\end{eqnarray}
They participate in the cancellation mechanism through the
 use of the relation (\ref{2.24}). We have
 constructed all possible terms of this type
 (for a few of the above structures there
 are actually no contributions), and found that the
 equations require that the corresponding coefficients vanish.
 Therefore these terms have no effect on our solutions, and we
 refrain from presenting their explicit parametrisation.
\par
This completes the discussion of all sectors
 of the Ansatz which contain
 the gravitino field and the gravitino curvature.
\vskip 8pt
\par

There are six sectors which contain the dilatino field
 $\lambda$.
 We constructed 21 independent terms of the structure
 $\bar\lambda\,\Gamma\psi_{(2)}R^3$:
\begin{eqnarray}
\label{L9}
 {\cal L}_{9}
 & = &
 +\{p_{1}R_{efgh}R_{abcd}R_{abcd}
 +p_{2}R_{efag}R_{bhcd}R_{abcd}
 +p_{3}R_{ghae}R_{bfcd}R_{abcd} \nonumber \\
&&\qquad
 +p_{4}R_{efab}R_{ghcd}R_{abcd}
 +p_{5}R_{egab}R_{fhcd}R_{abcd}
 +p_{6}R_{efab}R_{agcd}R_{bhcd} \nonumber \\
&&\qquad
 +p_{7}R_{egab}R_{afcd}R_{bhcd}
 +p_{8}R_{ghab}R_{aecd}R_{bfcd}
 +p_{9}R_{efac}R_{bgad}R_{bhcd} \nonumber \\
&&\qquad
 +p_{10}R_{egac}R_{bfad}R_{bhcd}
 +p_{11}R_{ghac}R_{bead}R_{bfcd}
 +p_{12}R_{aecg}R_{bfdh}R_{abcd} \nonumber \\
&&\qquad
 +p_{13}R_{aecg}R_{bfad}R_{bhcd}
 +p_{14}R_{aebg}R_{afcd}R_{bhcd}\}\,
 \bar\lambda\,\Gamma_{ef}\psi_{gh} \nonumber \\
&& +\{p_{15}R_{deah}R_{fgbc}R_{aibc}
 +p_{16}R_{hiad}R_{efbc}R_{agbc}
 +p_{17}R_{deah}R_{fibc}R_{agbc} \nonumber \\
&&\qquad
 +p_{18}R_{debh}R_{afci}R_{agbc}
 -p_{19}R_{deab}R_{fhac}R_{bgci}\}\,
 \bar\lambda\,\Gamma_{defg}\psi_{hi} \nonumber \\
&& +\{p_{20}R_{cdai}R_{efbj}R_{ghab}
 +p_{21}R_{cdij}R_{efab}R_{ghab}\}\,
 \bar\lambda\,\Gamma_{cdefgh}\psi_{ij}\,.
\end{eqnarray}
Besides the sector ${\cal L}_9$ there are the following
 $\lambda$-dependent contributions:
\begin{eqnarray}
\label{L10}
 {\cal L}_{10} & \sim & \bar\psi\lambda R^2{\cal D}R \,, \\
\label{L11}
 {\cal L}_{11} & \sim & \bar\psi{\cal D}\lambda R^3 \,, \\
\label{L12}
 {\cal L}_{12} & \sim & \bar\psi_{(2)}{\cal D}^2\lambda R^2 \,, \\
\label{L13}
 {\cal L}_{13} & \sim & \bar\psi_{(2)}{\cal D}\lambda R{\cal D}R
 \,, \\
\label{L14}
 {\cal L}_{14} & \sim & \bar\psi_{(2)}\lambda R{\cal D}^2 R \,, \\
\label{L15}
 {\cal L}_{15} & \sim & \bar\psi_{(2)}\lambda {\cal D}R {\cal D}R
 \,.
\end{eqnarray}
As we explained in Section 4,
 these additional sectors may be included,
 but are not actually required to achieve the cancellation
 of the variations we econsider.
 Since we
 we choose in Section 5 for the minimal option
 of including only
 (\ref{L9}) in the presentation our results,
 we shall not give the parametrization
 of ${\cal L}_{10-15}$ explicitly.
 For the same reason, we do not
 display $\lambda$-dependent terms containing
 more derivatives, which might participate through the use
 of (\ref{2.23}).

\newpage
\section{Appendix B}
This Appendix is devoted to the presentation of the two solutions we
 have found. If we choose in (\ref{5.1}) $b=0,\ b_1=b_2=b_3=0$
 and $a_1=1$ we obtain:
\begin{eqnarray}
\label{I1}
 e^{-1}{I}_1 & = &
 +R_{abcd}R_{abcd}R_{efgh}R_{efgh}
 -16R_{abcd}R_{abce}R_{dfgh}R_{efgh}\nonumber \\
&& +2R_{abcd}R_{abef}R_{cdgh}R_{efgh}
 +12R_{abce}R_{abdf}R_{cdgh}R_{efgh}\nonumber \\
&& -32R_{abce}R_{abdg}R_{cfdh}R_{efgh}
 +16R_{acbd}R_{aebg}R_{cfdh}R_{efgh}\nonumber \\
&& -16R_{acbe}R_{adbg}R_{cfdh}R_{efgh}\nonumber \\
&& \nonumber \\
&& +{\textstyle{{i}\over{8}}}\sqrt{2}\,e^{-1}
 \epsilon_{abcdefijkl}B_{ab}
 R_{cdgh}R_{efgh}R_{ijmn}R_{klmn}\nonumber \\
&& -{\textstyle{{i}\over{2}}}
 \sqrt{2}\,e^{-1}\epsilon_{abcdefhikl}
 B_{ab}R_{cdgj}R_{efgm}R_{hijn}R_{klmn}\nonumber \\
&& \nonumber \\
&& +\{8R_{bcad}{\cal D}_aR_{bcef}
 -8R_{abcd}{\cal D}_eR_{abcf}\}\,
 \bar\psi_{dg}\Gamma_{e}\psi_{fg}\nonumber \\
&& +\{-4R_{ceab}{\cal D}_dR_{fgab}
 -8R_{cfab}{\cal D}_dR_{egab}
 +16R_{acbf}{\cal D}_dR_{aebg} \nonumber \\
&&\qquad
 -16R_{aebf}{\cal D}_dR_{acbg}\}\,
 \bar\psi_{cd}\Gamma_{e}\psi_{fg}\nonumber \\
&& +4R_{deab}{\cal D}_cR_{fgab}
 \,\bar\psi_{ch}\Gamma_{def}\psi_{gh}\nonumber \\
&& +\{8R_{efac}{\cal D}_aR_{bdgh}
 +4R_{chaf}{\cal D}_aR_{bgde}
 -4R_{efac}{\cal D}_bR_{adgh}\}\,
 \bar\psi_{bc}\Gamma_{def}\psi_{gh}\nonumber \\
&& \nonumber \\
&& +16R_{acbd}R_{aebf}
 \,\bar\psi_{cg}\Gamma_{d}{\cal D}_e\psi_{fg}
 +32R_{bcaf}R_{deag}
 \,\bar\psi_{bc}\Gamma_{d}{\cal D}_e\psi_{fg}\nonumber \\
&& +8R_{cdaf}R_{egab}
 \,\bar\psi_{bh}\Gamma_{cde}{\cal D}_f\psi_{gh}\nonumber \\
&& \nonumber \\
&& +\{-20R_{afbg}R_{acde}R_{bcde}
 +20R_{afbc}R_{agde}R_{bcde} \nonumber \\
&&\qquad
 -32R_{bfad}R_{cgae}R_{bcde}\}\,
 \bar\psi_{h}\Gamma_{f}\psi_{gh}\nonumber \\
&& +\{-R_{efgh}R_{abcd}R_{abcd}
 +8R_{efag}R_{bhcd}R_{abcd}
 +4R_{ghae}R_{bfcd}R_{abcd} \nonumber \\
&&\qquad
 -4R_{ghaf}R_{becd}R_{abcd}
 -2R_{efab}R_{ghcd}R_{abcd}
 +2R_{egab}R_{fhcd}R_{abcd} \nonumber \\
&&\qquad
 +4R_{efab}R_{agcd}R_{bhcd}
 -28R_{egab}R_{afcd}R_{bhcd}
 +20R_{fgab}R_{aecd}R_{bhcd} \nonumber \\
&&\qquad
 +8R_{ghab}R_{aecd}R_{bfcd}
 +24R_{efac}R_{bgad}R_{bhcd}
 +48R_{egac}R_{bfad}R_{bhcd} \nonumber \\
&&\qquad
 +16R_{fgac}R_{bead}R_{bhcd}
 -16R_{ghac}R_{bead}R_{bfcd}
 -24R_{aecg}R_{bfdh}R_{abcd} \nonumber \\
&&\qquad
 +8R_{aecg}R_{bfad}R_{bhcd}
 +16R_{aebg}R_{afcd}R_{bhcd} \nonumber \\
&&\qquad -4R_{afbg}R_{aecd}R_{bhcd}\}\,
 \bar\psi_{e}\Gamma_{f}\psi_{gh}\nonumber \\
&& +\{{\textstyle{{1}\over{2}}}
 R_{efhi}R_{abcd}R_{abcd}
 -4R_{efah}R_{bicd}R_{abcd}
 -4R_{hiae}R_{bfcd}R_{abcd} \nonumber \\
&&\qquad
 +R_{efab}R_{hicd}R_{abcd}
 -2R_{efab}R_{ahcd}R_{bicd}
 +20R_{ehab}R_{afcd}R_{bicd} \nonumber \\
&&\qquad
 -2R_{hiab}R_{aecd}R_{bfcd}
 -16R_{ehac}R_{bfad}R_{bicd}
 -8R_{hiac}R_{bead}R_{bfcd} \nonumber \\
&&\qquad
 +8R_{aech}R_{bfdi}R_{abcd}
 -8R_{aech}R_{bfad}R_{bicd}\}\,
 \bar\psi_{g}\Gamma_{efg}\psi_{hi}\nonumber \\
&& +\{-8R_{efab}R_{agcd}R_{bhcd}
 -8R_{ehab}R_{afcd}R_{bgcd}
 +8R_{efac}R_{bgad}R_{bhcd} \nonumber \\
&&\qquad
 +4R_{ehac}R_{bfad}R_{bgcd}\}\,
 \bar\psi_{i}\Gamma_{efg}\psi_{hi}\nonumber \\
&& +\{-4R_{dhef}R_{agbc}R_{aibc}
 -2R_{efhi}R_{adbc}R_{agbc}
 -2R_{efad}R_{hibc}R_{agbc} \nonumber \\
&&\qquad
 +2R_{efah}R_{dibc}R_{agbc}
 -4R_{ehad}R_{fgbc}R_{aibc}
 -2R_{hiae}R_{fgbc}R_{adbc} \nonumber \\
&&\qquad
 +8R_{efbd}R_{agch}R_{aibc}
 -4R_{efbh}R_{adcg}R_{aibc}
 +8R_{efbh}R_{agci}R_{adbc} \nonumber \\
&&\qquad
 +4R_{efbh}R_{adci}R_{agbc}
 +16R_{debh}R_{aicf}R_{agbc}
 +16R_{ehbd}R_{aicf}R_{agbc} \nonumber \\
&&\qquad
 +8R_{hibe}R_{adcf}R_{agbc}
 -8R_{deab}R_{fgac}R_{hibc}
 -8R_{dhab}R_{efac}R_{gibc} \nonumber \\
&&\qquad
 -4R_{efab}R_{hiac}R_{bdcg}
 +8R_{ehab}R_{fiac}R_{bdcg}
 -12R_{efab}R_{ghac}R_{bdci} \nonumber \\
&&\qquad
 +4R_{efab}R_{dhac}R_{bgci}
 -16R_{deab}R_{fhac}R_{bgci}
 +8R_{ehab}R_{cdaf}R_{bgci} \nonumber \\
&&\qquad
 -8R_{adbe}R_{cfah}R_{bgci}\}\,
 \bar\psi_{d}\Gamma_{efg}\psi_{hi}\nonumber \\
&& +2R_{deab}R_{fgac}R_{bhci}
 \,\bar\psi_{j}\Gamma_{defgh}\psi_{ij}\nonumber \\
&& +\{2R_{deai}R_{fgbc}R_{ajbc}
 -12R_{deab}R_{fiac}R_{bgcj}\}\,
 \bar\psi_{h}\Gamma_{defgh}\psi_{ij}\nonumber \\
&& +\{-2R_{deci}R_{fgab}R_{hjab}
 -{\textstyle{{1}\over{2}}}
 R_{cdij}R_{efab}R_{ghab}
 +2R_{deac}R_{ijbf}R_{ghab} \nonumber \\
&&\qquad
 -4R_{diac}R_{efbj}R_{ghab}
 +4R_{deac}R_{fgbi}R_{hjab} \nonumber \\
&&\qquad
 -2R_{deac}R_{fgbi}R_{ahbj}\}\,
 \bar\psi_{c}\Gamma_{defgh}\psi_{ij}\nonumber \\
&& +\{-R_{cdaj}R_{efbk}R_{ghab}
 +{\textstyle{{1}\over{4}}}
 R_{cdjk}R_{efab}R_{ghab}\}\,
 \bar\psi_{i}\Gamma_{cdefghi}\psi_{jk}\nonumber \\
&& \nonumber \\
&& +\sqrt{2}\,\biggl[ \{
 R_{efgh}R_{abcd}R_{abcd}
 -8R_{efag}R_{bhcd}R_{abcd}
 -8R_{ghae}R_{bfcd}R_{abcd} \nonumber \\
&&\qquad\quad
 +2R_{efab}R_{ghcd}R_{abcd}
 -4R_{efab}R_{agcd}R_{bhcd}
 +40R_{egab}R_{afcd}R_{bhcd} \nonumber \\
&&\qquad\quad
 -4R_{ghab}R_{aecd}R_{bfcd}
 -32R_{egac}R_{bfad}R_{bhcd}
 +16R_{ghac}R_{bead}R_{bfcd} \nonumber \\
&&\qquad\quad
 +16R_{aecg}R_{bfdh}R_{abcd}
 -16R_{aecg}R_{bfad}R_{bhcd} \nonumber \\
&&\qquad\quad
 -32R_{aebg}R_{afcd}R_{bhcd}\}\,
 \bar\lambda\,\Gamma_{ef}\psi_{gh}\nonumber \\
&&\quad +\{-4R_{deah}R_{fgbc}R_{aibc}
 -24R_{deab}R_{fhac}R_{bgci}\}\,
 \bar\lambda\,\Gamma_{defg}\psi_{hi}\nonumber \\
&&\quad +\{-2R_{cdai}R_{efbj}R_{ghab}
 +{\textstyle{{1}\over{2}}}
 R_{cdij}R_{efab}R_{ghab}\}\,
 \bar\lambda\,\Gamma_{cdefgh}\psi_{ij}\biggr]\,.
\end{eqnarray}
\noindent
The modifications to the fermionic transformation rules follow
 from (\ref{5.5}). The result is:
\begin{eqnarray}
\delta_{\gamma}\psi_\mu & = & D_{b}\,\{(20 R_{bcde}R_{\mu
 cfg}R_{defg}
 -20 R_{bc\mu d}R_{cefg}R_{defg}\nonumber \\
&&\qquad\quad
 -32 R_{bdcf}R_{\mu ecg}R_{defg})\,\epsilon\}\nonumber \\
&&+D_{b}\,\{(2R_{\mu bcd}R_{efcd}R_{efgh}
 +4R_{\mu ecd}R_{bfcd}R_{efgh}
 -16R_{ce\mu d}R_{bcdf}R_{efgh} \nonumber \\
&&\qquad
 +24R_{\mu cde}R_{bcdf}R_{efgh}
 -24R_{\mu bcf}R_{cdeg}R_{defh}
 -12R_{\mu bcd}R_{cefg}R_{defh} \nonumber \\
&&\qquad
 -8R_{\mu fcd}R_{bceg}R_{defh}
 +8R_{cfde}R_{\mu bcg}R_{defh}
 -40R_{bdcf}R_{ce\mu g}R_{defh} \nonumber \\
&&\qquad
 +24R_{bfcd}R_{ce\mu g}R_{defh}
 -8R_{bfcd}R_{\mu ceg}R_{defh} \nonumber \\
&&\qquad
 +4R_{bcde}R_{cfde}R_{\mu fgh}
 -16R_{cedf}R_{bcdg}R_{\mu efh})\,
 \Gamma_{gh}\epsilon\}\nonumber \\
&& +D_{c}\,\{(-16R_{\mu dbf}R_{becg}R_{defh}
 -12R_{\mu bde}R_{bfcg}R_{defh})\,
 \Gamma_{gh}\epsilon\}\nonumber \\
&& +4D_{e}\,\{R_{\mu bcd}R_{bfcd}R_{efgh}
 \Gamma_{gh}\epsilon\}
 +D_{f}\,\{R_{bcde}R_{bcde}R_{\mu fgh}
 \Gamma_{gh}\epsilon\} \nonumber \\
&& +D_{g}\,\{(-28R_{\mu fbc}R_{bcde}R_{defh}
 +32R_{\mu bcd}R_{becf}R_{defh} \nonumber \\
&&\qquad
 -20R_{becd}R_{bfcd}R_{\mu efh})\,
 \Gamma_{gh}\epsilon\}\nonumber \\
&& +D_{b}\,\{(-2R_{\mu cde}R_{bcfg}R_{dehi}
 +2R_{bd\mu c}R_{cefg}R_{dehi} \nonumber \\
&&\qquad
 +2R_{bcde}R_{\mu cfg}R_{dehi}
 +12R_{bcdf}R_{ce\mu g}R_{dehi})\,
 \Gamma_{fghi}\epsilon\}\nonumber \\
&& +D_{c}\,\{(-8R_{\mu bdf}R_{becg}R_{dehi}
 +4R_{\mu dbf}R_{becg}R_{dehi})\,
 \Gamma_{fghi}\epsilon\}\nonumber \\
&& +D_{f}\,\{(-8R_{\mu cbd}R_{becg}R_{dehi}
 +8R_{\mu dbc}R_{bceg}R_{dehi} \nonumber \\
&&\qquad
 +4R_{bc\mu g}R_{bdeh}R_{cdei}
 -8R_{be\mu g}R_{cdbh}R_{cdei})\,
 \Gamma_{fghi}\epsilon\}\nonumber \\
&& +D_{b}\,\{({\textstyle{1\over 2}}
 R_{\mu bef}R_{cdgh}R_{cdij}
 -2R_{bcef}R_{\mu dgh}R_{cdij})\,
 \Gamma_{efghij}\epsilon\}\nonumber \\
&& -2D_{e}\,\{R_{\mu bcf}R_{bdgh}R_{cdij}
 \Gamma_{efghij}\epsilon\}\nonumber \\
&& \nonumber \\
\label{trans1}
 \delta_{\gamma}\lambda& = &
 -{\textstyle{1\over 4}}
 \sqrt{2}\,\Gamma^\mu\delta_{\gamma}\psi_\mu\,.
\end{eqnarray}
Note that $\delta_\gamma\psi$ contains
 $R^3 {\cal D}\epsilon$-terms. The appearance of new
 supersymmetry transformations containing ${\cal D}\epsilon$
 can easily be avoided. The contributions of the
 equation of motion $\Psi$ in (\ref{5.5}) are, schematically,
 $R^3\bar\epsilon\,{\cal D}\Psi$, or, after a partial
 integration:
\begin{eqnarray}
\label{Beps1}
 -({\cal D}R^3)\bar\epsilon\,\Psi
 -R^3({\cal D}\bar\epsilon)\Psi \,.
\end{eqnarray}
The first term must be cancelled by changing
 the transformation rule of the gravitino.
 The second term can also be cancelled by adding to the action:
\begin{eqnarray}
\label{Bnew}
 R^3 \bar\psi\,\Psi \,.
\end{eqnarray}
Of course the new term has to varied. The variation of $\psi$
 gives ${\cal D}\epsilon$ and cancels the second term in
 (\ref{Beps1}) (this time we do not perform the partial integration
 away from $\epsilon$!). The variation of $\Psi$ gives a
 combination of bosonic equations of motion, and this can be
 cancelled
 by changing the bosonic transformation rules. If this
 procedure is followed, the new fermionic transformation rules are
 as in (\ref{trans1}), but without the ${\cal D}\epsilon$-terms.
\vskip 8pt
The second solution is obtained by taking in (\ref{5.1}) $a_1=1,\
b=8,\ b_1=b_2=0,\ b_3=-48\sqrt{2}$:
\begin{eqnarray}
\label{I2}
 e^{-1} I_2 & = &
 +R_{abcd}R_{abcd}R_{efgh}R_{efgh}
 -8R_{abcd}R_{abce}R_{dfgh}R_{efgh}\nonumber \\
&& +2R_{abcd}R_{abef}R_{cdgh}R_{efgh}
 -4R_{abce}R_{abdf}R_{cdgh}R_{efgh}\nonumber \\
&& \nonumber \\
&& -48\sqrt{2}\,H_{abd}R_{abce}R_{ghcf}{\cal D}_dR_{efgh}
 +96\sqrt{2}\,H_{abe}R_{abcd}R_{ghcf}{\cal D}_dR_{efgh}\nonumber \\
&& \nonumber \\
&& +{\textstyle{{i}\over{8}}}\sqrt{2}\,e^{-1}
 \epsilon_{abcdefijkl}
 B_{ab}R_{cdgh}R_{efgh}R_{ijmn}R_{klmn}\nonumber \\
&& \nonumber \\
&& +4R_{cdab}{\cal D}_eR_{fgab}
 \,\bar\psi_{cd}\Gamma_{e}\psi_{fg}
 -2R_{bcaf}{\cal D}_aR_{bcde}
 \,\bar\psi_{gh}\Gamma_{def}\psi_{gh}\nonumber \\
&& -4R_{ghaf}{\cal D}_aR_{bcde}
 \,\bar\psi_{bc}\Gamma_{def}\psi_{gh}\nonumber \\
&& \nonumber \\
&& -8R_{adbc}R_{aebc}
 \,\bar\psi_{fg}\Gamma_{d}{\cal D}_e\psi_{fg}
 -16R_{bcad}R_{fgae}
 \,\bar\psi_{bc}\Gamma_{d}{\cal D}_e\psi_{fg} \nonumber \\
&& \nonumber \\
&& +\{-R_{efgh}R_{abcd}R_{abcd}
 +16R_{egaf}R_{bhcd}R_{abcd}
 +12R_{ghae}R_{bfcd}R_{abcd} \nonumber \\
&&\qquad
 -4R_{ghaf}R_{becd}R_{abcd}
 -2R_{efab}R_{ghcd}R_{abcd}
 +4R_{efab}R_{agcd}R_{bhcd} \nonumber \\
&&\qquad
 +8R_{egab}R_{afcd}R_{bhcd}
 +8R_{fgab}R_{aecd}R_{bhcd} \nonumber \\
&&\qquad
 +8R_{ghab}R_{aecd}R_{bfcd}\}\,
 \bar\psi_{e}\Gamma_{f}\psi_{gh}\nonumber \\
&& +\{{\textstyle{{1}\over{2}}}
 R_{efhi}R_{abcd}R_{abcd}
 -4R_{hiae}R_{bfcd}R_{abcd}
 +R_{efab}R_{hicd}R_{abcd} \nonumber \\
&&\qquad
 -2R_{hiab}R_{aecd}R_{bfcd}\}\,
 \bar\psi_{g}\Gamma_{efg}\psi_{hi}\nonumber \\
&& +\{-4R_{efhi}R_{adbc}R_{agbc}
 -4R_{hiae}R_{fgbc}R_{adbc} \nonumber \\
&&\qquad
 -4R_{hiad}R_{efbc}R_{agbc}\}\,
 \bar\psi_{d}\Gamma_{efg}\psi_{hi} \nonumber \\
&& +\{-{\textstyle{{1}\over{2}}}
 R_{cdij}R_{efab}R_{ghab}
 -R_{deij}R_{cfab}R_{ghab} \}\,
 \bar\psi_{c}\Gamma_{defgh}\psi_{ij}\nonumber \\
&& +{\textstyle{{1}\over{4}}}
 R_{cdjk}R_{efab}R_{ghab}
 \,\bar\psi_{i}\Gamma_{cdefghi}\psi_{jk}\nonumber \\
&& \nonumber \\
&& +\sqrt{2}\,\biggl[\{
 R_{efgh}R_{abcd}R_{abcd}
 -8R_{ghae}R_{bfcd}R_{abcd}
 +2R_{efab}R_{ghcd}R_{abcd} \nonumber \\
&&\qquad\quad
 -4R_{ghab}R_{aecd}R_{bfcd}\}\,
 \bar\lambda\,\Gamma_{ef}\psi_{gh}\nonumber \\
&&\quad +{\textstyle{{1}\over{2}}}
 R_{cdij}R_{efab}R_{ghab}
 \,\bar\lambda\,\Gamma_{cdefgh}\psi_{ij}\biggr]\,.
\end{eqnarray}
The modifications to the transformation rules
 can be calculated from (\ref{5.6}). We find:
\begin{eqnarray}
\delta_{\gamma}\psi_\mu & = & D_{b}\,\{(2R_{\mu bcd}R_{cdef}R_{efgh}
 \Gamma_{gh}
 -4R_{\mu bcf}R_{decg}R_{defh})
 \Gamma_{gh}\epsilon\}\nonumber \\
&& -8D_{e}\,\{R_{bcdf}R_{bcdg}R_{\mu efh}
 \Gamma_{gh}\epsilon\}
 +D_{f}\,\{R_{bcde}R_{bcde}R_{\mu fgh}
 \Gamma_{gh}\epsilon\}\nonumber \\
&& +{\textstyle{1\over 2}}
 D_{b}\,\{R_{\mu bef}R_{cdgh}R_{cdij}
 \Gamma_{efghij}\epsilon\}\nonumber \\
&& \nonumber \\
\label{trans2}
 \delta_{\gamma}\lambda & = &
 -{\textstyle{1\over 4}}\sqrt{2}\,
 \Gamma^\mu\delta_{\gamma}\psi_\mu\,.
\end{eqnarray}

\vskip 8pt
The solution $I_2$ has a Yang-Mills analogon. The proper way to
 derive
this Yang-Mills solution from $I_2$ consists in two steps. First,
by using pair exchange (\ref{pairex}), the
$R^4$-terms must be written
in such a way that the contraction over Lorentz indices corresponds
to the Yang-Mills trace.
Second, the spin connection must be written with $H$-torsion. These
steps do not require the use of the identities (\ref{B-Identities}).
We use the notation $W_{\mu\nu}= {\rm tr}\,F_{\mu\nu}\chi$.
The result is:
\begin{eqnarray}
\label{LYM}
 e^{-1}I_{YM} & = &-{\textstyle{1\over 2}}t^{\mu_1...\mu_8}
 \,{\rm tr}\,F_{\mu_1\mu_2}F_{\mu_3\mu_4}
 \,{\rm tr}\,F_{\mu_5\mu_6}F_{\mu_7\mu_8} \nonumber \\
&&+{\textstyle{i\over 8}}\sqrt{2}\,e^{-1}
 \epsilon^{\mu_1....\mu_{10}}B_{\mu_1\mu_2}
 \,{\rm tr}\,F_{\mu_3\mu_4}F_{\mu_5\mu_6}
 \,{\rm tr}\,F_{\mu_7\mu_8}F_{\mu_9\mu_{10}} \nonumber \\
&& \nonumber \\
&&+4\,\bar W^{\mu\nu}\Gamma^\lambda\,
 {\rm tr}\,\chi\,{\cal D}_\lambda F_{\mu\nu}
 -2\,{\rm tr}\,F^{\mu\nu}{\cal D}_\mu F^{\lambda\rho}\,
 {\rm tr}\,\bar\chi\,\Gamma_{\nu\lambda\rho}\chi \nonumber \\
&&-4\,\bar W^{\mu\nu}\Gamma^{\lambda\rho}{}_\nu\,{\rm tr}\,\chi\,
 {\cal D}_\mu F_{\lambda\rho} \nonumber \\
&& \nonumber \\
&&-8\,{\rm tr}\,F^{\mu\lambda}F^\nu{}_\lambda
 \,{\rm tr}\,\bar\chi\,\Gamma_\mu{\cal D}_\nu\chi
 -16\,\bar W^{\mu\nu}\Gamma_\nu
 \,{\rm tr}\,({\cal D}^\lambda\chi)F_{\mu\lambda}
 \nonumber \\
&& \nonumber \\
&& +\{-\,{\rm tr}\,F^{\mu\nu}F_{\mu\nu}
 \bar\psi_{\lambda}\Gamma_{\rho}
 -12\,{\rm tr}\,F^{\mu\nu}F_{\mu\lambda}
 \bar\psi_{\rho}\Gamma_{\nu}
 +12\,{\rm tr}\,F^{\mu\nu}F_{\mu\lambda}
 \bar\psi_{\nu}\Gamma_{\rho} \nonumber \\
&&\qquad
 -2\,{\rm tr}\,F^{\mu\nu}F_{\lambda\rho}
 \bar\psi_{\mu}\Gamma_{\nu}
 +12\,{\rm tr}\,F^{\mu}{}_{\lambda}F^{\nu}{}_{\rho}
 \bar\psi_{\mu}\Gamma_{\nu}\}\,W^{\lambda\rho}
 \nonumber \\
&& +\{{\textstyle{1\over 2}}
 \,{\rm tr}\,F^{\mu\nu}F_{\mu\nu}
 \bar\psi_{\sigma}\Gamma^{\sigma}{}_{\lambda\rho}
 -4\,{\rm tr}\,F^{\mu\nu}F_{\mu\lambda}
 \bar\psi_{\sigma}\Gamma^{\sigma}{}_{\nu\rho}
 +\,{\rm tr}\,F^{\mu\nu}F_{\lambda\rho}
 \bar\psi_{\sigma}\Gamma^{\sigma}{}_{\mu\nu} \nonumber \\
&&\qquad
 -2\,{\rm tr}\,F^{\mu}{}_{\lambda}F^{\nu}{}_{\rho}
 \bar\psi_{\sigma}\Gamma^{\sigma}{}_{\mu\nu}
 -4\,{\rm tr}\,F^{\mu\nu}F_{\mu}{}^{\sigma}
 \bar\psi_{\nu}\Gamma_{\sigma\lambda\rho}
 +4\,{\rm tr}\,F^{\mu}{}_{\lambda}F^{\nu\sigma}
 \bar\psi_{\mu}\Gamma_{\nu\sigma\rho} \nonumber \\
&&\qquad
 +4\,{\rm tr}\,F^{\mu\nu}F^{\sigma}{}_{\lambda}
 \bar\psi_{\rho}\Gamma_{\mu\nu\sigma} \}\,W^{\lambda\rho}
 \nonumber \\
&& +\{-{\textstyle{1\over 2}}
 \,{\rm tr}\,F^{\mu\nu}F^{\sigma\tau}
 \bar\psi_{\lambda}\Gamma_{\mu\nu\sigma\tau\rho}
 -\,{\rm tr}\,F^{\mu\nu}F^{\sigma\tau}
 \bar\psi_{\mu}\Gamma_{\nu\sigma\tau\lambda\rho}\}\,
 W^{\lambda\rho} \nonumber \\
&& +{\textstyle{1\over 4}}
 \,{\rm tr}\,F^{\mu\nu}F^{\sigma\tau}
 \bar\psi_{\xi}\Gamma^{\xi}{}_{\mu\nu\sigma\tau\lambda\rho}
 W^{\lambda\rho} \nonumber \\
&& \nonumber \\
&& +\sqrt{2}\,\biggl[
 \{ \,{\rm tr}\,F^{\mu\nu}F_{\mu\nu}
 \,\bar\lambda\,\Gamma_{\lambda\rho}
 -8 \,{\rm tr}\,F^{\mu\nu}F_{\mu\lambda}
 \,\bar\lambda\,\Gamma_{\nu\rho}
 -4 \,{\rm tr}\,F^{\mu}{}_{\lambda}F_{\nu\rho}
 \,\bar\lambda\,\Gamma_{\mu\nu} \nonumber \\
&&\qquad\quad
 +2 \,{\rm tr}\,F^{\mu\nu}F_{\lambda\rho}
 \,\bar\lambda\,\Gamma_{\mu\nu} \}\,W^{\lambda\rho}
 \nonumber \\
&&\quad +{\textstyle{1\over 2}}
 \,{\rm tr}\,F^{\mu\nu}F^{\sigma\tau}
 \,\bar\lambda\,\Gamma_{\mu\nu\sigma\tau\lambda\rho}\,
 W^{\lambda\rho} \biggr] \,.
\end{eqnarray}
The additional supersymmetry transformation rules of $\chi$ follow
 from (\ref{5.7}), and read:
\begin{eqnarray}
 \label{trans3}
 \delta_\gamma\chi & = &
 -{\textstyle{1\over 4}}
 \Gamma_{cdefgh}\epsilon\, F_{gh}\,{\rm tr}\,F_{cd}F_{ef}
 \nonumber \\
 && +\Gamma_{ef}\epsilon\, F_{cd}
 \,(2\,{\rm tr}\,F_{ce}F_{df} -\,{\rm tr}\,F_{cd}F_{ef})
 \nonumber \\
 && \qquad
 - {\textstyle{1\over 2}} \Gamma_{ef}\epsilon\, F_{ef}
 \,{\rm tr}\, F_{cd}F_{cd}
 + 4\Gamma_{ef}\epsilon\, F_{df}
 \,{\rm tr}\,F_{cd}F_{ce} \,.
\end{eqnarray}

\end{document}